\documentclass[%
 reprint,
 amsmath,amssymb,
 aps,
 prab,
floatfix,
]{revtex4-2}

\usepackage{textcomp}
\usepackage[T1]{fontenc}
\usepackage{lmodern}
\usepackage{tgtermes}
\usepackage{newtxmath}
\usepackage{siunitx}
\usepackage{booktabs} 
\usepackage{placeins}
\usepackage{verbatim}
\usepackage[most]{tcolorbox}

\usepackage{graphicx}
\usepackage{dcolumn}
\usepackage{bm}

\newcommand{\eps}[1]{\varepsilon_{#1}}
\newcommand{\F}[1]{\phi_{#1}}
\newcommand{\cH}[1]{\mathcal{H}_{#1}}
\newcommand\clog{C_\text{log}}

\begin{document}

\preprint{APS/123-QED}

\title{Interplay of space charge and intrabeam scattering in the LHC ion injector chain}

\author{M.~Zampetakis}
\email{michail.zampetakis@cern.ch}
\affiliation{European Organization for Nuclear Research (CERN), CH-1211 Geneva 23, Switzerland;}
\affiliation{Department of Physics, University of Crete, P.O. Box 2208, GR-71003 Heraklion, Greece}
\author{F.~Antoniou}
\author{F.~Asvesta}
\author{H.~Bartosik}
\author{Y.~Papaphilippou}
\affiliation{European Organization for Nuclear Research (CERN), CH-1211 Geneva 23, Switzerland;}
\author{A.~Sa\'a~Hern\'andez}
\affiliation{Instituto Galego de Física de Altas Enerxías (IGFAE), Universidade de Santiago de Compostela, Santiago de Compostela, Spain}

\begin{abstract}
    The ion injectors of the CERN accelerator chain, in particular the Super Proton Synchrotron~(SPS) and the Low Energy Ion Ring~(LEIR), operate in a strong space charge~(SC) and intrabeam scattering~(IBS) regime, which can degrade beam quality. Optimizing the ion beam performance therefore requires studying the interplay of these two effects in tracking simulations by incorporating both SC and IBS effects interleaved with lattice nonlinearities. In this respect, the kinetic theory approach of treating IBS effects has been deployed. A modified stochastic IBS kick is introduced, with friction and diffusion coefficients derived in two forms: an efficient Nagaitsev formulation based on complete elliptic integrals of the second kind, and a complete Bjorken--Mtingwa formulation. This IBS kick is implemented in PyORBIT and extensively benchmarked against analytical models. Results of combined space charge and intrabeam scattering simulations for the SPS and LEIR are presented and compared with observations from beam measurements.
\end{abstract}

\maketitle

\section{\label{sec:1}Introduction}
Incoherent effects like Intrabeam Scattering~(IBS) and Space Charge~(SC) can significantly limit the beam performance of hadron synchrotrons and have therefore been intensively studied in several accelerators and regimes. In particular, IBS plays an important role in ion and proton storage rings, in which the beam is stored for many hours~\cite{MSc_Mertens, JWei_RHIC, Fischer_RHIC, Bruce:ibs_kick, Lebedev_1, Papadopoulou:2018uvl, PhD_Papadopoulou}, in damping rings, light sources and linear accelerators~\cite{Lebedev_2, PhD_Antoniou, Antoniou:SLS, Bane_2,Bane_KEK, Kubo_KEK, Huang, Xiao_2010, DiMitri_2020} and in very low energy ion machines~\cite{Hunt_Carli,RHIC,Siggel}. On the other hand, SC effects have been studied in many low-energy machines~\cite{Franchetti:2003aa, Metral:2006qw, Franchetti:2010zz, Franchetti:2017aa, Asvesta:ps, SaaHernandez:2018zqv, Fermi_res, Fermi_IOTA, Fermi_boost, Fermi_highI, SC_RHIC, SC_eRHIC} and damping rings~\cite{Venturini:2006rp, Venturini:2006ilc, Venturini:2007ler, Xiao:2007ilc}, as the SC-induced tune shift can result in particle losses and transverse emittance increase due to periodic resonance crossing. 

The interplay between IBS and SC can further enhance particle diffusion in phase space in the presence of excited resonances, as was shown in simulation studies for the Compact LInear Collider Damping Rings~(CLIC DRs)~\cite{zampetakis:clic}. In rings that operate below the transition energy, IBS can lead to emittance exchange between the longitudinal and the transverse planes. This mechanism could not be simulated with the simplified IBS kick implemented in the tracking simulations described in Refs.~\cite{zampetakis:clic, Bruce:ibs_kick}. Thus, a more general implementation of the IBS kick is required to take into account the exchange of momenta in all planes.

The most commonly used numerical method for simulating the three-dimensional IBS effects is the ``binary collisions'' model~(BCM). In this model, a distribution of macroparticles is divided into cells and each macroparticle is considered to interact with any other neighboring macroparticle during a specific time window. The probability of this interaction depends on the macroparticle density in each specific cell. The BCM model is implemented in various tracking codes, e.g.~the MOnte-CArlo Code~(MOCAC)~\cite{mocac}, the Software for IBS and Radiation Effects~(SIRE)~\cite{Vivoli:1} and the IBStrack~\cite{osti_1} found in the collective effects simulation tool CMAD~\cite{Pivi:1, Sonnad:1}. Some of these codes include other effects like electron cooling and radiation damping as well.

The downside of the BCM model is that a large number of macroparticles is required (typically $>5\times 10^{4}$) in order to have sufficiently populated cells. In the case of turn-by-turn calculations in the presence of additional dynamical effects, this translates to a high demand on computational resources. To overcome this challenge, a three-dimensional approximate model was introduced in the past~\cite{Zenkevich:0, Zenkevich:1, Zenkevich:2}, arising from the general kinetic theory of gases. This model is based on the Fokker--Planck partial differential equation~(FPE) that describes the evolution over time of the probability density function of the velocity of a particle that experiences friction and random forces~\cite{Risken1996}.

To progress in the understanding of the performance limitations of the heavy-ion injector chain at CERN, it is important to have a reliable tool to simulate IBS together with other effects. Such a simulation tool will also be crucial for studies of future collider projects. In~\cite{zampetakis:clic}, a simple kick was used to simulate the IBS effect in tracking simulations for rings that operate above transition where no equilibrium can exist~\cite{Piwinski:1974it}. This means that all three planes grow indefinitely in the presence of IBS alone, if no other limitations are exceeded. However, for rings operating below transition, equilibrium among the three planes can exist. Thus, simulating the IBS effect entails the additional complexity that the momenta exchange due to IBS can lead to damping of the emittance in one or more planes~\cite{Piwinski:1974it}. In this case, a simple diffusive kick is not sufficient and a friction force is needed to simulate the particles' momenta exchange. In this paper, a general stochastic IBS kick is introduced, based on the kinetic description of IBS~\cite{Zenkevich:0, Zenkevich:1, Zenkevich:2}, with the effective kick coefficients evaluated using Nagaitsev's formalism~\cite{Nagaitsev}. This implementation includes both friction and diffusion terms and is therefore capable of treating rings operating below transition, where momentum exchange between the transverse and longitudinal planes can lead to damping in one or more planes. Using this kick, a simulation campaign was performed for the CERN ion injectors, in particular the Low Energy Ion Ring~(LEIR) and the Super Proton Synchrotron~(SPS). The goal of these simulations is to understand and recreate experimental observations of transverse emittance growth, that so far could not be explained by stand-alone IBS and SC simulations, by studying the interplay between SC and IBS.

This paper is organized in the following way. After this Introduction, Sec.~\ref{sec:3} introduces the kinetic description of IBS and the modified stochastic IBS kick used in the tracking simulations. The kick includes both friction and diffusion terms, allowing the exchange of momentum between the transverse and longitudinal planes to be modeled. The evaluation of the effective kick coefficients using Nagaitsev's formalism~\cite{Nagaitsev}, based on complete elliptic integrals of the second kind, is also presented in Sec.~\ref{sec:3}. For completeness, the Bjorken--Mtingwa formalism~\cite{Bjorken:0}, the original approximate kinetic model, and the corresponding BM-based derivation of the effective kick coefficients are summarized in Appendix~\ref{app:BM_kinetic}. In Sec.~\ref{sec:4}, the simulation setup is discussed and a thorough benchmarking of the introduced IBS kick is shown for different configurations in LEIR and the SPS. Section~\ref{sec:5} summarizes the LEIR study case. First, the operational aspects of LEIR are discussed. Then, stand-alone tracking simulations with SC and IBS, as well as combined SC and IBS simulations, are shown. Similarly, in Sec.~\ref{sec:6}, a case for the SPS is presented, where beam measurements are compared with tracking simulations in the presence of the two effects. A dedicated Discussion is presented in Sec.~\ref{sec:7}, where the physical mechanism behind the interplay between space charge and intrabeam scattering is qualitatively analyzed and interpreted based on the simulation results. The main conclusions of this work are presented in Sec.~\ref{sec:8}.

\section{\label{sec:3}Kinetic description of IBS}
Following the kinetic approach of Refs.~\cite{Zenkevich:0, Zenkevich:1, Zenkevich:2}, the cumulative effect of IBS can be described as a combination of friction and diffusion in momentum space. The friction term accounts for the systematic exchange of momentum between the transverse and longitudinal degrees of freedom, while the diffusion term represents the stochastic nature of multiple small-angle Coulomb collisions.

In the convention of the original approximate model, the position and dimensionless momentum variables are defined as
\begin{align}
    \vec{r} =
        \begin{pmatrix}
            z-z_s \\           
            x \\
            y
        \end{pmatrix},\quad\quad
    \vec{p} =
        \begin{pmatrix}
            \frac{1}{\gamma}\frac{\Delta p}{p_0} \\           
            x' \\
            y'
        \end{pmatrix},
\end{align}
and the corresponding evolution of the distribution function can be written in the form
\begin{equation}\label{eq:FPE}
    \frac{\partial \Phi}{\partial t}
    =
    -\frac{\partial}{\partial p_m} (F_m\Phi)
    +\frac{1}{2}
    \frac{\partial^2}{\partial p_m \partial p_{m'}}
    (D_{m,m'}\Phi),
\end{equation}
where $z$ is the longitudinal particle coordinate, $z_s$ is the longitudinal coordinate of the bunch center, $x$ and $y$ are the horizontal and vertical transverse coordinates, $p_0$ is the reference-particle momentum, $\Delta p=p-p_0$ is the particle's momentum deviation, $x'=p_x/p_0$, $y'=p_y/p_0$, with $p_x$ and $p_y$ being the horizontal and vertical momentum components. Furthermore, $\gamma$ is the Lorentz factor, $\Phi$ is the distribution function, $F_m$ are the components of the friction force, and $D_{m,m'}$ are the diffusion coefficients.

In the original approximate model, the friction force is assumed to be linear in momentum, $F_m=-K_m p_m$, while the diffusion coefficients are taken as constants. This leads to a Langevin equation of the form
\begin{equation}\label{eq:LE}
    p_i(t+\Delta t)
    =
    p_i(t)-K_i p_i(t)\Delta t
    +\sqrt{\Delta t}\sum_{j=1}^3 C_{i,j}\varsigma_j,
\end{equation}
where $\varsigma_j$ are normally distributed random numbers and the matrix $C_{i,j}$ is related to the diffusion tensor. The relation between the diffusion coefficients and the matrix $C_{i,j}$ follows the procedure described in Ref.~\cite{Zenkevich:0}; the corresponding friction and diffusion coefficients of the original approximate model are recalled in Appendix~\ref{app:kinetic}.

In the present work, this kinetic description is used as the basis for a modified stochastic IBS kick. Instead of applying the original approximate-model coefficients directly, we derive effective friction and diffusion coefficients from the analytical IBS growth rates. These coefficients are evaluated using the formalism of Nagaitsev~\cite{Nagaitsev}, as described in the following section.

\subsection{Modified approximate IBS model and stochastic kick}\label{sec:3a}
The original approximate model was benchmarked successfully against the binary-collision model for the TWAC storage ring~\cite{Zenkevich:0}. However, applying the same approximate model~\cite{Zenkevich:0,Zenkevich:2} to the CERN ion injectors did not reproduce the behavior expected from analytical IBS models such as Bjorken--Mtingwa~\cite{Bjorken:0} or Nagaitsev~\cite{Nagaitsev}, especially in the horizontal plane.

Inspired by the structure of the approximate model, we introduce a modified approximate IBS model. The goal is to retain the kinetic description in terms of friction and diffusion, while deriving effective kick coefficients that reproduce the analytical IBS growth rates.

Following the same simplification steps used in the Bjorken--Mtingwa formalism, the friction and diffusion coefficients can be written as
\begin{equation}
    K_i^{\mathrm{BM}}=\frac{A_\mathrm{BM}}{\left< p_i^2 \right>} 
    \int_{0}^{\infty} d\lambda\,
    \frac{\lambda^{1/2}}{\sqrt{\det \mathscr{L}}}
    \mathscr{L}^{-1}_{i,i},
    \label{eq:K1}
\end{equation}
\begin{equation}
    D_{i,j}^{\mathrm{BM}}=A_\mathrm{BM}
    \int_{0}^{\infty} d\lambda\,
    \frac{\lambda^{1/2}}{\sqrt{\det \mathscr{L}}}
    \left(\delta_{ij}\mathrm{Tr}\mathscr{L}^{-1}
    -\mathscr{L}^{-1}_{i,j}\right),
    \label{eq:D1}
\end{equation}
where $\mathscr{L}=L+\lambda I$, with $I$ the unit matrix. The definitions of the auxiliary matrix $L$ and of the Bjorken--Mtingwa diffusion kernels are given in Appendix~\ref{app:BM}. In this form, the diffusion kernels can be related to the friction and diffusion coefficients through
\begin{equation}\label{eq:IKD}
    I_{ij}^{\mathrm{BM}} =
    \begin{cases}
        D_{i,i}^{\mathrm{BM}}
        -2\left\langle p_i^2\right\rangle K_i^{\mathrm{BM}},
        & i=j, \\[1mm]
        3D_{i,j}^{\mathrm{BM}},
        & i\neq j .
    \end{cases}
\end{equation}

For the off-diagonal elements, the factor of three follows directly from comparing Eq.~\eqref{eq:D1} with the corresponding Bjorken--Mtingwa diffusion kernels defined in Appendix~\ref{app:BM}. No off-diagonal friction coefficient is introduced. Equation~\eqref{eq:IKD} provides the link between the kinetic formulation and the analytical IBS growth rates. Substituting this relation into the analytical growth-rate expressions separates the diffusion and friction contributions and directly defines the effective plane-dependent coefficients $G_u$ and $F_u$, with $u=x,y,z$, that enter the stochastic kick. With these coefficients, the kick produces the same emittance and momentum-spread growth rates as the analytical IBS model. The explicit expressions obtained using the Bjorken--Mtingwa formalism, including vertical dispersion, are given in Appendix~\ref{app:BM_modified}. In the simulations presented in this work, the same kick structure is used, but the coefficients are evaluated using Nagaitsev's formalism, as described in Sec.~\ref{AM_nag}.

To apply the IBS kick to the momenta of each macroparticle in the distribution, for each plane $u=x,y,z$, a modified Langevin equation is used. This equation is inspired by the simple IBS kick initially introduced to incorporate the longitudinal line density~\cite{Blaskiewicz:2007, Blaskiewicz:2008}, which was later used in IBS studies~\cite{Bruce:ibs_kick, zampetakis:clic}. In the present model, the same idea is generalized to include both friction and diffusion terms in all three planes:
\begin{equation}\label{eq:mod_LE}
    p_u(t+\Delta t) = p_u(t) - F_u p^c_u(t)\rho(z)\Delta t
    + \sigma_{p_u} \sqrt{2 G_u \Delta t \rho(z)} \varsigma_u,
\end{equation}
where $\varsigma_u$ is a random number following a normal distribution with zero mean and unit standard deviation. The factor $\rho(z)=2\sigma_z\sqrt{\pi}\lambda(z)$ accounts for the local longitudinal line density, where $\lambda(z)$ is the normalized longitudinal line density at position $z$ within the bunch, $\sigma_z$ is the bunch length, and $\sigma_{p_u}$ is the RMS momentum in plane $u$. In the implementation, the momentum entering the friction term is evaluated relative to the bunch centroid, i.e., $p^c_u = p_u-\langle p_u\rangle$. 

Here, $\Delta t=T_{\mathrm{rev}}$, as one effective IBS kick is applied per revolution. Since IBS acts on timescales much longer than the revolution period and does not rely on resonance behavior, this approximation reproduces the slow emittance evolution while maintaining computational efficiency. The model is intended for cases where the turn-by-turn emittance change remains small.

For the transverse planes, the momenta $p_{x,y}$ are transformed by subtracting the dispersive contribution and using the local optics at the position of the kick:
\begin{equation}\label{eq:norm_p}
    \tilde{p}_u^i = \left(p_u^i-\eta'_u\frac{\Delta p_i}{p_0}\right) 
    \times \sqrt{\beta_u} + 
    \left( u_i - \eta_u\frac{\Delta p_i}{p_0}\right)
    \times 
    \frac{\alpha_u}{\sqrt{\beta_u}}, \quad u=x,y.
\end{equation}
Here, $\beta_u$ and $\alpha_u$ are the Twiss parameters, while $\eta_u$ and $\eta'_u$ are the dispersion function and its derivative in plane $u$.

After applying the kick, the transverse momenta are converted back to the physical reference system, including the corresponding off-momentum contribution.

\subsection{Evaluation of the kick coefficients using the complete elliptic integrals of the second kind}
\label{AM_nag}
The modified kick of Eq.~\eqref{eq:mod_LE} requires the effective friction and diffusion coefficients $F_u$ and $G_u$. In the simulations presented in this work, these coefficients are evaluated using the formalism of Nagaitsev~\cite{Nagaitsev}, which expresses the IBS integrals through complete elliptic integrals of the second kind. This approach is computationally efficient and therefore suitable for long macroparticle tracking simulations.

It should be noted that, in its standard form, Nagaitsev's formalism neglects vertical dispersion. The more general derivation using the Bjorken--Mtingwa formalism, which includes vertical dispersion, is given in Appendix~\ref{app:BM_modified}. In the simulations presented here, the Nagaitsev-based expressions are used for the evaluation of the kick coefficients.

Following Carlson's definition, the complete symmetric elliptic integral of the second kind is given by
\begin{equation}
    R_D(x,y,z)=\frac{3}{2}\int_{0}^{\infty} 
    \frac{dt}{\sqrt{(t+x)(t+y)(t+z)^3}},
\end{equation}
and can be evaluated efficiently using algorithms based on the duplication theorem~\cite{Carlson}.

Following Nagaitsev~\cite{Nagaitsev}, the IBS growth rates can be expressed through the three integrals
\begin{equation}\label{eq:l1}
    R_1=\frac{1}{\lambda_1}R_D\Big(\frac{1}{\lambda_2},\frac{1}{\lambda_3},\frac{1}{\lambda_1}\Big),
\end{equation}
\begin{equation}\label{eq:l2}
    R_2=\frac{1}{\lambda_2}R_D\Big(\frac{1}{\lambda_3},\frac{1}{\lambda_1},\frac{1}{\lambda_2}\Big),
\end{equation}
\begin{equation}\label{eq:l3}
    R_3=\frac{1}{\lambda_3}R_D\Big(\frac{1}{\lambda_1},\frac{1}{\lambda_2},\frac{1}{\lambda_3}\Big),
\end{equation}
where $\lambda_1,\lambda_2,\lambda_3$ are the eigenvalues of the auxiliary matrix used in Nagaitsev's formalism. They are written as
\begin{equation}\label{eq:eig1}
    \lambda_1=a_y,
\end{equation}
\begin{equation}\label{eq:eig2}
    \lambda_2=a_1+q_x,
\end{equation}
\begin{equation}\label{eq:eig3}
    \lambda_3=a_1-q_x,
\end{equation}
with
\begin{equation}\label{eq:nag1}
    a_x=\frac{\beta_x}{\eps{x}}, \quad a_y=\frac{\beta_y}{\eps{y}}, \quad
    a_s=a_x \Big( \frac{\eta_x^2}{\beta_x^2}+\F{x}^2 \Big)+\frac{1}{\sigma_{\delta}^2},
\end{equation}
\begin{equation}
    a_1=\frac{1}{2}(a_x+\gamma^2a_s), \quad 
    a_2=\frac{1}{2}(a_x-\gamma^2a_s),
\end{equation}
\begin{equation}\label{eq:nag2}
    q_x=\sqrt{a_2^2+\gamma^2a_x^2\F{x}^2}.
\end{equation}

Here, the optics quantity $\F{x}$ is defined as
\begin{equation}
    \F{x}=\eta'_x-\frac{\beta'_x\eta_x}{2\beta_x},
\end{equation}
and the corresponding dispersion invariant is
\begin{equation}
    \cH{x}=\frac{1}{\beta_x}\left[
    \eta_x^2+
    \left(\beta_x\eta'_x-\frac{1}{2}\beta'_x\eta_x\right)^2
    \right].
\end{equation}

To connect the elliptic integrals to the friction and diffusion coefficients entering the stochastic kick, we introduce the integrals
\begin{equation}\label{eq:Thij}
    \Theta_{i,j} =
    \int_{0}^{\infty} d\lambda\,
    \frac{\lambda^{1/2}}{\sqrt{\det \mathscr{L}}}
    \mathscr{L}^{-1}_{i,j},
\end{equation}
where $\mathscr{L}=L+\lambda I$, with $I$ the unit matrix and $L$ the auxiliary matrix of the Bjorken--Mtingwa formalism, recalled in Appendix~\ref{app:BM}. In the Nagaitsev formalism, the relevant $\Theta_{i,j}$ are expressed in terms of $R_1$, $R_2$, and $R_3$.

The integrals $\Theta_{i,j}$ are introduced to express the friction and diffusion coefficients $K_i$ and $D_{i,j}$ entering the stochastic kick in the Nagaitsev formalism. Matching the resulting growth-rate combinations to Nagaitsev's partial growth rates~\cite{Nagaitsev} gives the relations below:
\begin{equation}
    \Theta_{x,x} + \Theta_{z,z} = R_2 + R_3, \qquad
    \Theta_{y,y} = R_1,
    \label{eq:Ndfy}
\end{equation}
\begin{equation}
    \Theta_{x,x} = \frac{1}{2}
        \left[
        R_2 \left(1+\frac{a_2}{q_x}\right)
        + R_3 \left(1-\frac{a_2}{q_x}\right)
        \right],
\end{equation}
\begin{equation}
    \Theta_{z,z} = \frac{1}{2}
        \left[
        R_2 \left(1-\frac{a_2}{q_x}\right)
        + R_3 \left(1+\frac{a_2}{q_x}\right)
        \right],
\end{equation}
\begin{equation}
    \Theta_{x,z} = \frac{\gamma\F{x}a_x}{2q_x} \left(R_3-R_2\right).
\end{equation}

Using these expressions, the Nagaitsev-based diffusion and friction terms entering the modified kick can be written directly in terms of the elliptic integrals as
\begin{equation}\label{eq:Dxx}
    D_{x,x} = \frac{1}{2}\left[
                2R_1
                + R_2\left(1-\frac{a_2}{q_x}\right)
                + R_3\left(1+\frac{a_2}{q_x}\right) \right],
\end{equation}
\begin{equation}
    D_{z,z} = \frac{\gamma^2}{2}\left[
                2R_1
                + R_2\left(1+\frac{a_2}{q_x}\right)
                + R_3\left(1-\frac{a_2}{q_x}\right) \right],
\end{equation}
\begin{equation}
    D_{x,z} = 6\gamma\F{x}\Theta_{x,z} 
            = \frac{3\gamma^2\F{x}^2a_x}{q_x} \left(R_3-R_2\right),
\end{equation}
\begin{equation}
    K_x = R_2\left(1+\frac{a_2}{q_x}\right)
        +R_3\left(1-\frac{a_2}{q_x}\right),
\end{equation}
\begin{equation}
    K_z = \gamma^2\left[
        R_2\left(1-\frac{a_2}{q_x}\right)
        +R_3\left(1+\frac{a_2}{q_x}\right)
        \right],
\end{equation}
\begin{equation}\label{eq:Dyy}
    D_{y,y}=R_2+R_3, \quad K_y=2R_1.
\end{equation}

The quantities $D_{i,j}$ and $K_i$ are intermediate combinations of the elliptic integrals. They are combined below to obtain the effective coefficients $G_{x,y,z}$ and $F_{x,y,z}$, chosen such that $T_{x,y,z}^{-1}=G_{x,y,z}-F_{x,y,z}$. These coefficients determine, respectively, the stochastic and deterministic terms in Eq.~\eqref{eq:mod_LE}. For a lattice element of length $L_e$, these coefficients are
\begin{equation}\label{nag:Gx}
    \begin{aligned}
        G_x={}&\frac{1}{\eps{x}}
        \frac{Nr_0^2c\clog}{12\pi\beta^3\gamma^5\sigma_z}
        \\
        &\times\int^{L_e}_0
        \frac{\beta_x\,ds}{L_e\sigma_x\sigma_y}
        \Bigg[D_{x,x}+
        \Bigg(\frac{\eta_x^2}{\beta_x^2}+\F{x}^2\Bigg)D_{z,z}
        +D_{x,z}\Bigg],
    \end{aligned}
\end{equation}
\begin{equation}
    G_y=\frac{1}{\eps{y}}\frac{Nr_0^2c\clog}{12\pi\beta^3\gamma^5\sigma_z}
    \int^{L_e}_0 \frac{\beta_y ds}{L_e\sigma_x\sigma_y}D_{y,y},
\end{equation}
\begin{equation}
    G_z=\frac{1}{\sigma_{\delta}^2}\frac{Nr_0^2c\clog}{12\pi\beta^3\gamma^5\sigma_z}
    \int^{L_e}_0 \frac{ds}{L_e\sigma_x\sigma_y} D_{z,z},
\end{equation}
\begin{equation}
    \begin{aligned}
        F_x={}&\frac{1}{\eps{x}}
        \frac{Nr_0^2c\clog}{12\pi\beta^3\gamma^5\sigma_z}
        \\
        &\times\int^{L_e}_0
        \frac{\beta_x\,ds}{L_e\sigma_x\sigma_y}
        \Bigg[K_x+
        \Bigg(\frac{\eta_x^2}{\beta_x^2}+\F{x}^2\Bigg)K_z\Bigg],
    \end{aligned}
\end{equation}
\begin{equation}
    F_y=\frac{1}{\eps{y}}\frac{Nr_0^2c\clog}{12\pi\beta^3\gamma^5\sigma_z}
    \int^{L_e}_0 \frac{\beta_y\,ds}{L_e\sigma_x\sigma_y}K_y,
\end{equation}
\begin{equation}\label{nag:Fz}
    F_z=\frac{1}{\sigma_{\delta}^2}\frac{Nr_0^2c\clog}{12\pi\beta^3\gamma^5\sigma_z}
    \int^{L_e}_0 \frac{ds}{L_e\sigma_x\sigma_y} K_z.
\end{equation}
Here, $\eps{x,y}$ are the geometric RMS transverse emittances, $\sigma_{x,y}$ are the local RMS transverse beam sizes, $\sigma_\delta$ is the RMS relative momentum spread, $\beta=v/c$ is the relativistic velocity factor, $c$ is the speed of light in vacuum, $N$ denotes the total number of physical particles per bunch, $r_0$ is the classical radius of the corresponding particle species, and $C_{\mathrm{log}}$ is the Coulomb logarithm.

The expressions for $G_{x,y,z}$ and $F_{x,y,z}$ represent averages over individual lattice elements of length $L_e$. The effective coefficients used in tracking are obtained by taking the length-weighted average of these element-wise values around the ring circumference $C$. The coefficients $G_i$ and $F_i$ are therefore the Nagaitsev-based coefficients used in the stochastic kick of Eq.~\eqref{eq:mod_LE}. To simulate IBS for a coasting beam, the bunch length has to be replaced by $\sigma_z=C/(2\sqrt{\pi})$ in these expressions, and one has to consider that $T_z^\mathrm{coasting}=2T_z^\mathrm{bunched}$~\cite{Piwinski:1974it}.

For completeness, the corresponding construction of the effective coefficients using the Bjorken--Mtingwa formalism is provided in Appendix~\ref{app:BM_modified}.

\subsection{Practical Implementation of the IBS Kick Model}\label{IBSrecipe}
The analytical formulation introduced in the previous subsections is implemented in the macroparticle tracking simulations as an effective stochastic kick model. This subsection summarizes the numerical procedure used to compute and apply the IBS-induced kicks.

In the simulations presented in this work, one effective IBS kick is applied to each macroparticle once per turn using Eq.~\eqref{eq:mod_LE}. For the transverse planes, the momenta are first transformed according to Eq.~\eqref{eq:norm_p}, so that the dispersive contribution is removed and the kick is applied in normalized phase-space variables. After the kick, the momenta are transformed back to the physical reference system.

The effective coefficients entering the kick, $G_i$ and $F_i$ with $i=x,y,z$, are evaluated using the Nagaitsev formalism described in Sec.~\ref{AM_nag}. The diffusion and friction coefficients $D_{i,j}$ and $K_i$ are first computed from Eqs.~(\ref{eq:Dxx}--\ref{eq:Dyy}), using the optical functions of Eqs.~(\ref{eq:nag1}--\ref{eq:nag2}) and the elliptic integrals $R_{1,2,3}$ defined in Eqs.~(\ref{eq:l1}--\ref{eq:l3}). These quantities are then combined to obtain the effective kick coefficients $G_i$ and $F_i$ from Eqs.~(\ref{nag:Gx}--\ref{nag:Fz}).

The coefficients $G_i$ and $F_i$ are evaluated for each lattice element of length $L_e$ and averaged around the ring to obtain the effective coefficients used for the turn-by-turn kick. Since they depend on the instantaneous beam parameters, namely the transverse emittances, bunch length, and momentum spread, they are re-evaluated periodically during the simulation. In the studies presented here, the update interval was 500~turns for LEIR and 1000~turns for the SPS.

In Eq.~\eqref{eq:mod_LE}, the deterministic term proportional to $F_i$ represents the friction contribution, while the stochastic term proportional to $G_i$ represents the diffusion contribution. The random variable is sampled from a Gaussian distribution with zero mean and unit standard deviation. When averaged over many turns and particles, this procedure reproduces the analytical IBS growth rates while allowing the effect to be included in macroparticle tracking simulations together with other collective effects.

For clarity, the simulation procedure can be summarized as follows:
\begin{enumerate}
    \item Evaluate the lattice functions and the instantaneous beam parameters.
    \item Compute the diffusion and friction coefficients $D_{i,j}$ and $K_i$ from Eqs.~(\ref{eq:Dxx}--\ref{eq:Dyy}).
    \item Compute the effective kick coefficients $G_i$ and $F_i$ from Eqs.~(\ref{nag:Gx}--\ref{nag:Fz}).
    \item Transform the transverse momenta according to Eq.~\eqref{eq:norm_p}.
    \item Apply the stochastic and deterministic IBS kick to each macroparticle using Eq.~\eqref{eq:mod_LE}.
    \item Transform the transverse momenta back to the physical reference system and continue tracking through the ring.
    \item Repeat the procedure, updating the coefficients periodically as the beam parameters evolve.
\end{enumerate}

This formulation provides a self-consistent numerical implementation of the Nagaitsev formalism within the macroparticle tracking framework. The same stochastic kick structure can also be formulated using the Bjorken--Mtingwa formalism; the corresponding expressions for the effective coefficients are given in Appendix~\ref{app:BM_modified}.

\section{\label{sec:4}Simulation set-up and benchmarking of the IBS model}
The simulation studies of the incoherent effects of SC and IBS were performed using the Polymorphic Tracking Code~(PTC)~\cite{PTC} in the PyORBIT~\cite{pyorbit} simulation library. PyORBIT is a well-known simulation tool that has been intensively benchmarked and used for SC studies for many different accelerators~\cite{shishlo:pyorbit, Asvesta:ps, PhD_Asvesta, Schmidt:2016izr, Li:pyorbit, Yuan:pyorbit}. It provides the possibility to use various models for the SC potential, e.g.~a fully self-consistent 2.5D Particle-In-Cell~(PIC) solver, a slice-by-slice SC solver and an analytical frozen potential solver. 

For the presented studies, the SC effect is included using the SC frozen potential solver, where the SC kick at a given position in the ring is analytically calculated from the lattice functions, the beam intensity and the transverse beam sizes using the Bassetti-Erskine formula~\cite{Bassetti}. This space charge kick is weighted by the line density at the longitudinal position of the particle being tracked. The kick is applied at multiple locations around the ring, known as SC nodes. A convergence study for the number of SC nodes used in the presented simulations can be found in Appendix~\ref{app:conv}.

The term ``frozen potential'' in SC simulations refers to a model where the electrostatic potential generated by a bunch of charged particles is assumed to be static or ``frozen'', meaning it does not evolve as the beam parameters evolve. In studies that also account for excited resonances or IBS, the assumptions underlying the ``frozen'' space charge potential may no longer hold. In order to achieve a quasi-self-consistent evaluation of the SC kick during the simulation run, the SC potential was recomputed every 500~turns in LEIR and 1000~turns in the SPS, according to the evolution of the transverse beam parameters and the longitudinal line density. This method is known as the ``adaptive frozen model'', which allows for a more accurate representation of dynamic effects while reducing noise effects from the limited number of macroparticles.

In order to perform the studies of the interplay between the incoherent effects of SC and IBS, an additional IBS module was implemented inside PyORBIT. The IBS kick used in the tracking simulations is constructed from the Nagaitsev-based coefficients described in Sec.~\ref{AM_nag}. The purpose of the benchmarking presented here is not to validate the analytical formalism itself, but to verify that the continuous IBS growth rates are correctly translated into discrete stochastic kicks applied to the macroparticles. In other words, the benchmark demonstrates that the implemented kick reproduces the macroscopic emittance evolution predicted by the analytical IBS model.

The module was benchmarked against analytical calculations for LEIR and the SPS. In this paper, we present benchmark cases for Pb ions in different parameter regimes in LEIR and the SPS. The LEIR benchmarking cases are presented first. In these cases, the IBS kick is applied once per turn, while the IBS coefficients are re-evaluated every $500$~turns according to the evolution of the transverse beam parameters and the longitudinal line density. Each simulation is tracked for $250000$~turns, corresponding to approximately $0.69$~s at the nominal LEIR injection energy. The initial macroparticle distribution is Gaussian in the transverse planes. To reproduce the experimental observations, the initial longitudinal distribution is generated as a binomial distribution $g(x)=(1-x^2)^\mu$ with exponent $\mu=15$, which is not far from a Gaussian distribution. In the parameter tables below, $N_c$ denotes the total number of elementary charges per bunch. The physical particle number $N$ used in the IBS calculations is obtained by dividing $N_c$ by the charge state of the corresponding ion.

The benchmark cases presented below were chosen to test the IBS kick in different regimes. Since LEIR and the SPS operate below transition in the nominal cases considered here, IBS can lead to an exchange between the longitudinal and transverse degrees of freedom. As a result, depending on the initial beam parameters, the transverse emittances may grow while the momentum spread decreases, increases, or remains approximately constant. These different behaviors are determined by the relative IBS growth rates in the three planes and by the distance of the initial beam parameters from the IBS equilibrium.

\begin{figure}[t]
    \centering
    \includegraphics*[trim=18 10 17 15, clip, width=1.\columnwidth]{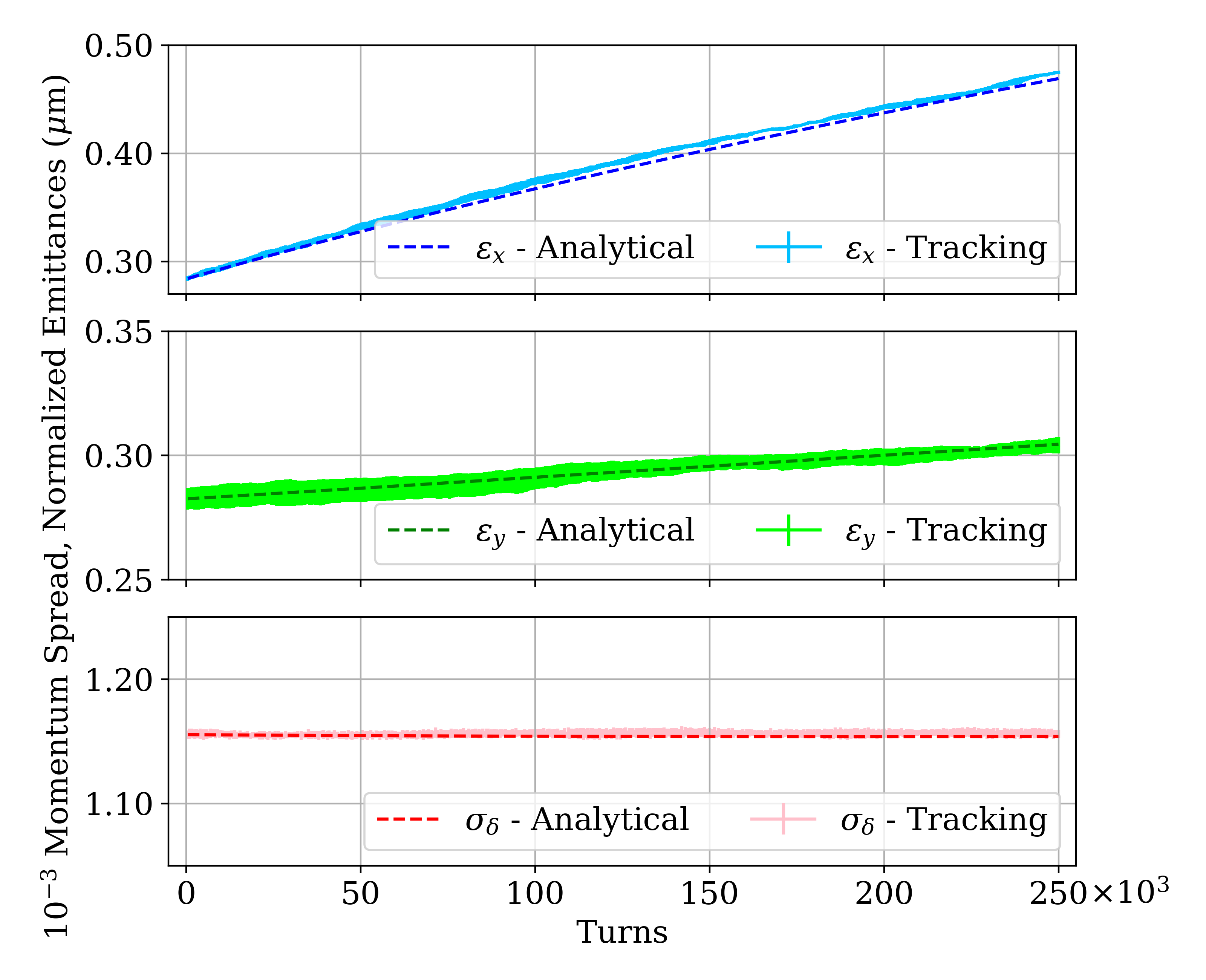}
    \caption{Comparison of the time evolution of the horizontal~(top) and vertical~(middle) emittances, and the momentum spread~(bottom), between analytical IBS predictions~(dark dashed lines) and tracking simulations including IBS~(solid lines), for the nominal LEIR parameters listed in Table~\ref{tab:LEIR_params}. Error bars indicate the standard deviation over three independent simulations.}
   \label{fig:bench_nom}
\end{figure}

Figure~\ref{fig:bench_nom} shows the evolution of the horizontal normalized emittance~(top), vertical normalized emittance~(middle), and momentum spread~(bottom), comparing tracking simulations using the modified approximate IBS model~(solid lines) with the analytical predictions based on Nagaitsev's formalism~(dark dashed lines) for the nominal parameters of LEIR, as described in Table~\ref{tab:LEIR_params}. The transverse emittances and the momentum spread are evaluated using the second-order moments of the macroparticle distribution, following an approach similar to that described in~\cite{Franchetti:2017aa}. In particular, a Gaussian fit is applied to the beam profile, and then the weighted RMS emittances are calculated from the actual beam profile up to the $8\sigma$ extent of the Gaussian fit. Excellent agreement is observed in all three planes. For the nominal LEIR parameters, the transverse emittances grow slowly, while the momentum spread remains almost constant. This indicates that, for these initial parameters, the longitudinal plane is close to the IBS equilibrium condition, whereas the transverse planes still experience moderate IBS-driven emittance growth.
\begin{table}[!h]
   \centering
   \caption{The LEIR Parameters}
   \begin{tabular}{lc}
       \toprule
       \multicolumn{2}{c}{\textbf{Ring and Beam Parameters}}     \\
       \midrule
        Circumference [m]                & 78.54         \\
        $Pb^{54+}$ rest energy [GeV]     & 193.7         \\
        Injection Kinetic energy [GeV/u] & 0.0042        \\
        Relativistic gamma, $\gamma_r$   & 1.0045        \\
        Gamma transition, $\gamma_\textrm{tr}$  & 2.8384        \\
        Harmonic number, $h$             & 2             \\
        RF voltage [MV]                  & 0.0011        \\
        Momentum compaction factor       & 0.1241        \\
        \midrule
        Total number of charges, $N_c$   & 2.1$\times10^{10}$    \\
        Normalized horizontal emittance, $\varepsilon_x^N$ [$\mu$m] & 0.282    \\
        Normalized vertical emittance, $\varepsilon_y^N$ [$\mu$m]   & 0.282    \\
        Bunch length, $\sigma_z$[m]                    & 4.2     \\
       \bottomrule
   \end{tabular}
   \label{tab:LEIR_params}
\end{table}

\begin{figure}[t]
    \centering
    \includegraphics*[trim=18 10 17 15, clip, width=1.\columnwidth]{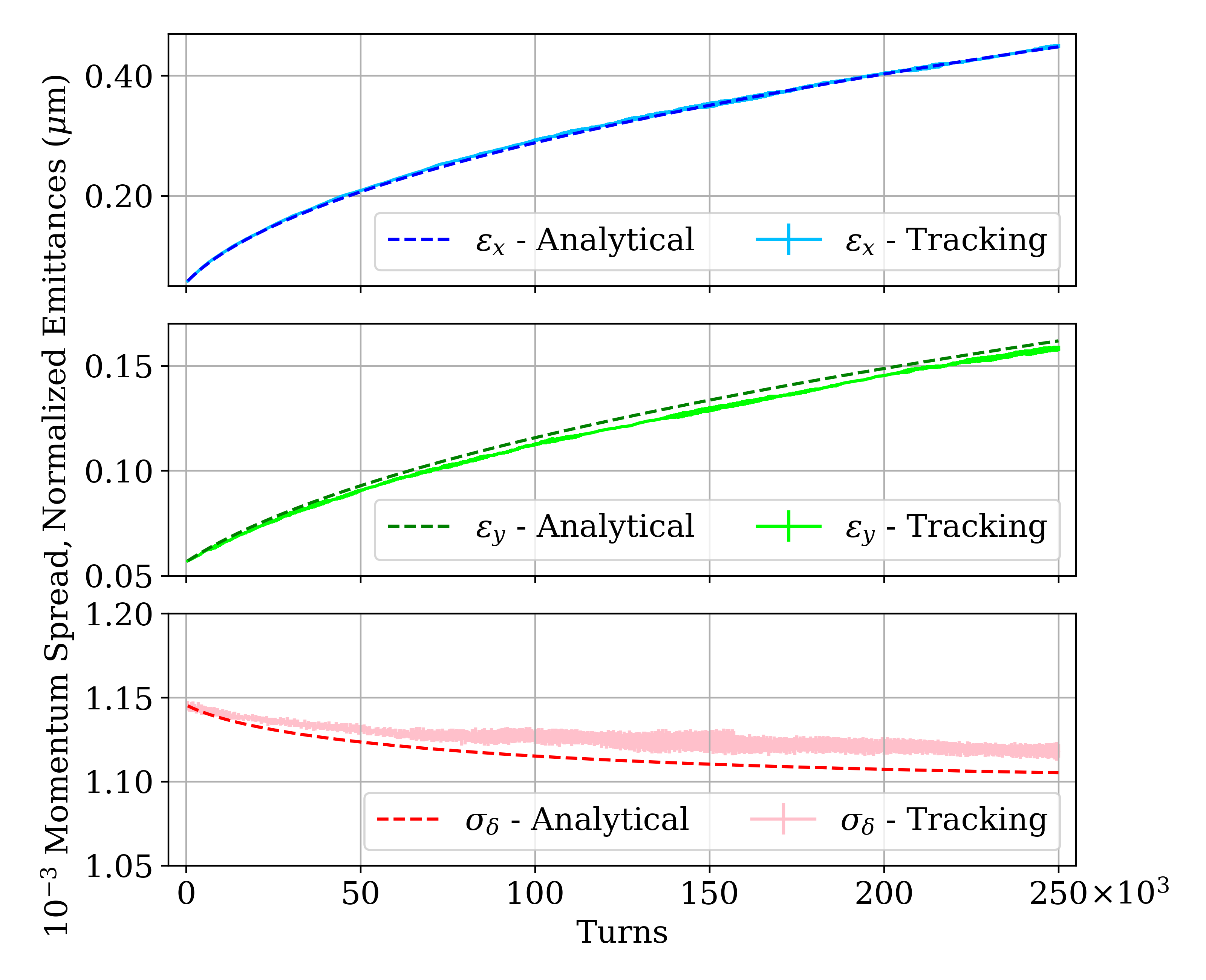}
    \caption{Comparison of the time evolution of the horizontal~(top) and vertical~(middle) emittances, and the momentum spread~(bottom), between analytical IBS predictions~(dark dashed lines) and tracking simulations including IBS~(solid lines), for a round beam with $\varepsilon_{x,y}^N=0.056~\mu$m and $\sigma_z=4.2$~m in LEIR. Error bars indicate the standard deviation over three independent simulations.}
    \label{fig:bench_ediv5}
\end{figure}

To further benchmark the IBS kick, another benchmarking case was performed in the regime of a round beam with small normalized transverse emittances of $\varepsilon_{x,y}^N=0.056~\mu$m. The rest of the parameters remain the nominal ones. A similar comparison between the tracking simulation including the IBS kick and the analytical estimations is shown in Fig.~\ref{fig:bench_ediv5}. As expected, IBS becomes stronger in this case because of the smaller transverse emittances and therefore the larger beam density. More blow-up is produced in the horizontal plane, while the momentum spread is damped. This behavior is expected below transition, where IBS can drive an exchange between the longitudinal and transverse degrees of freedom. In this case, the beam evolves toward equilibrium by transferring part of the longitudinal momentum spread into transverse emittance growth. Good agreement is observed again between the tracking simulations and the analytical predictions.

\begin{figure}[!ht]
    \centering
    \includegraphics*[trim=15 15 15 15, clip, width=1.\columnwidth]{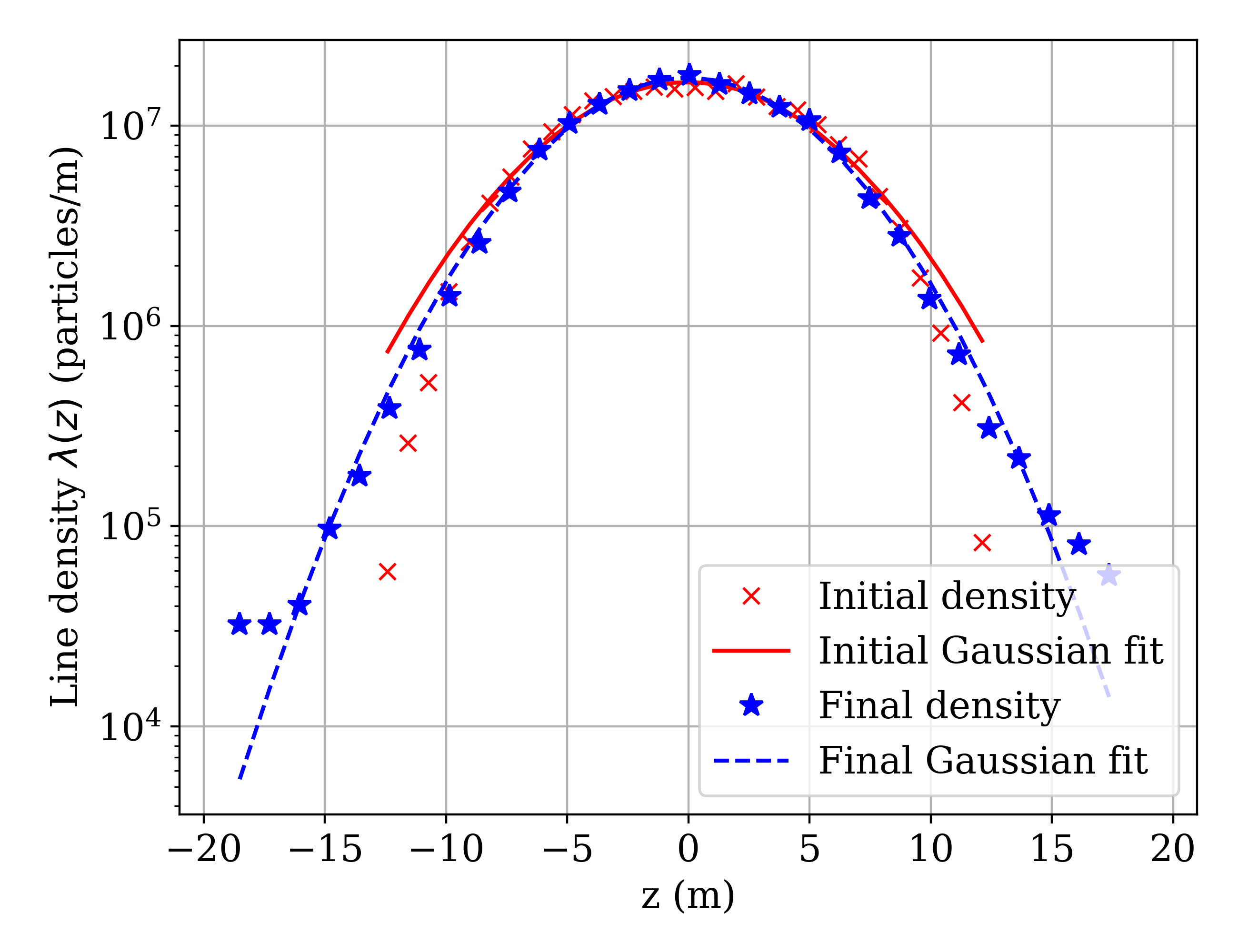}
    \caption{Initial~(red) and final~(blue) longitudinal line density of the bunch from tracking simulations with IBS for a round beam with $\varepsilon_{x,y}^N=0.056~\mu$m and a bunch length of $\sigma_z=4.2$~m. Crosses indicate the simulated profiles, while the curves show Gaussian fits. The initial distribution is binomial; after IBS, the core becomes closer to Gaussian but the tails are more populated, highlighting the non-Gaussian features responsible for small differences with analytical predictions.}
    \label{fig:z_profiles_LEIR}
\end{figure}

The largest discrepancy appears in the longitudinal plane. The analytical model assumes Gaussian distributions, whereas in the tracking simulations the initial distribution is binomial. As the IBS kicks act, the core of the distribution becomes more Gaussian, but the tails remain more populated, as illustrated in Fig.~\ref{fig:z_profiles_LEIR}. The figure shows the initial~(red) and final~(blue) longitudinal line density of the bunch~(markers), compared with Gaussian fits~(curves). This deviation from a perfect Gaussian, particularly in the tails, can explain the small differences observed with respect to the analytical predictions.

\begin{figure}[h]
    \centering
    \includegraphics*[trim=18 10 17 15, clip, width=1.\columnwidth]{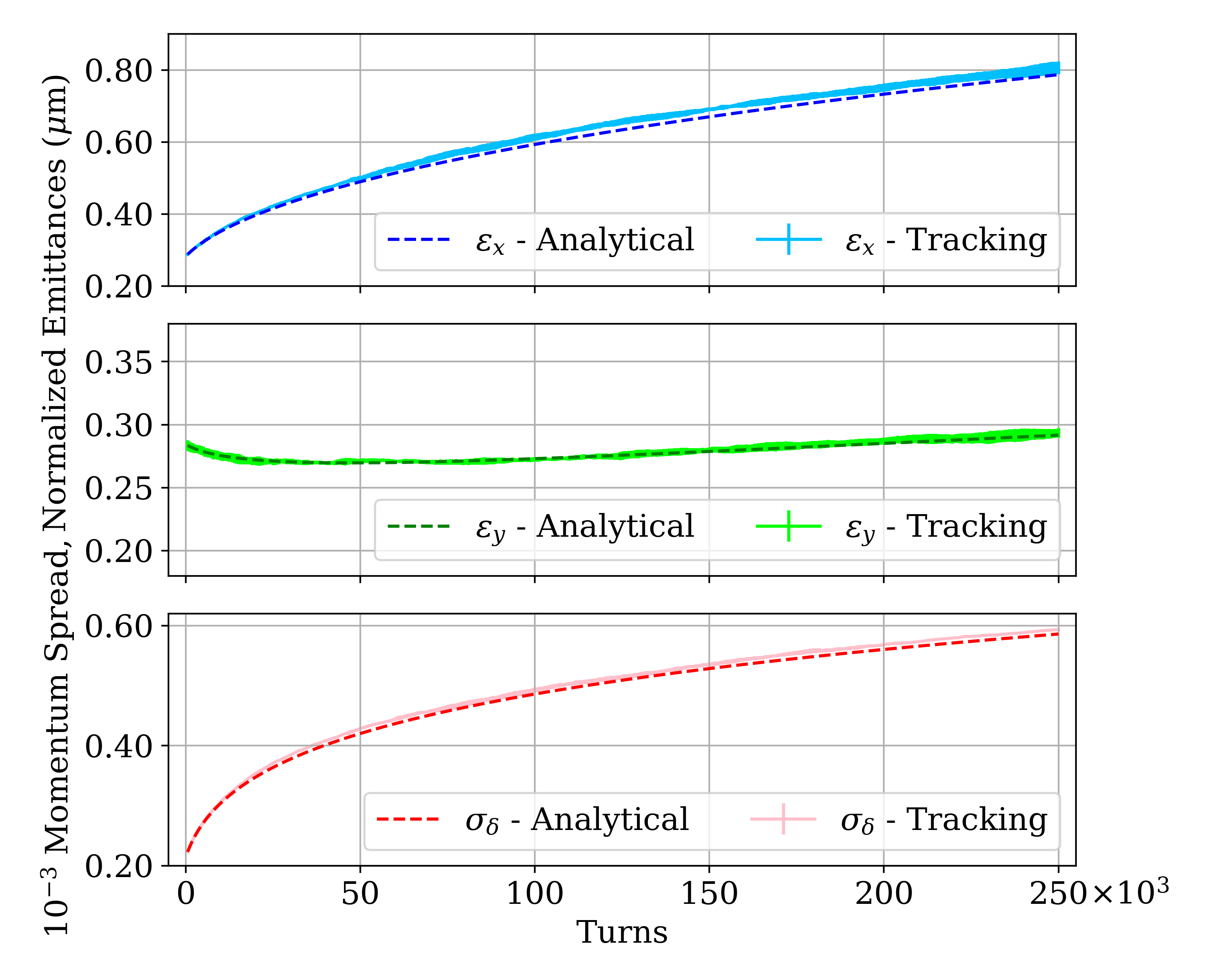}
    \caption{Comparison of the time evolution of the horizontal~(top) and vertical~(middle) emittances, and the momentum spread~(bottom), between analytical IBS predictions~(dark dashed lines) and tracking simulations including IBS~(solid lines), for a round beam with $\varepsilon_{x,y}^N=0.282~\mu$m and $\sigma_z=0.9$~m in the LEIR. Error bars indicate the standard deviation over three independent simulations.}
   \label{fig:bench_londiv5}
\end{figure}

Figure~\ref{fig:bench_londiv5} shows a comparison for the short-bunch regime. In this case, the momentum spread and bunch length are approximately one fifth of their operational values, while the beam is transversely round with the nominal normalized emittances $\varepsilon_{x,y}^N=0.282~\mu$m. The horizontal emittance and momentum spread grow strongly, whereas the vertical emittance initially decreases and subsequently increases. For these initial conditions, the small initial momentum spread produces a positive longitudinal IBS growth rate. The vertical IBS growth rate is initially negative but changes sign as the beam parameters evolve toward the IBS equilibrium. Very good agreement is again observed between the tracking simulations and the analytical calculations.

\begin{figure}[h]
    \centering
    \includegraphics*[trim=18 10 17 15, clip, width=1.\columnwidth]{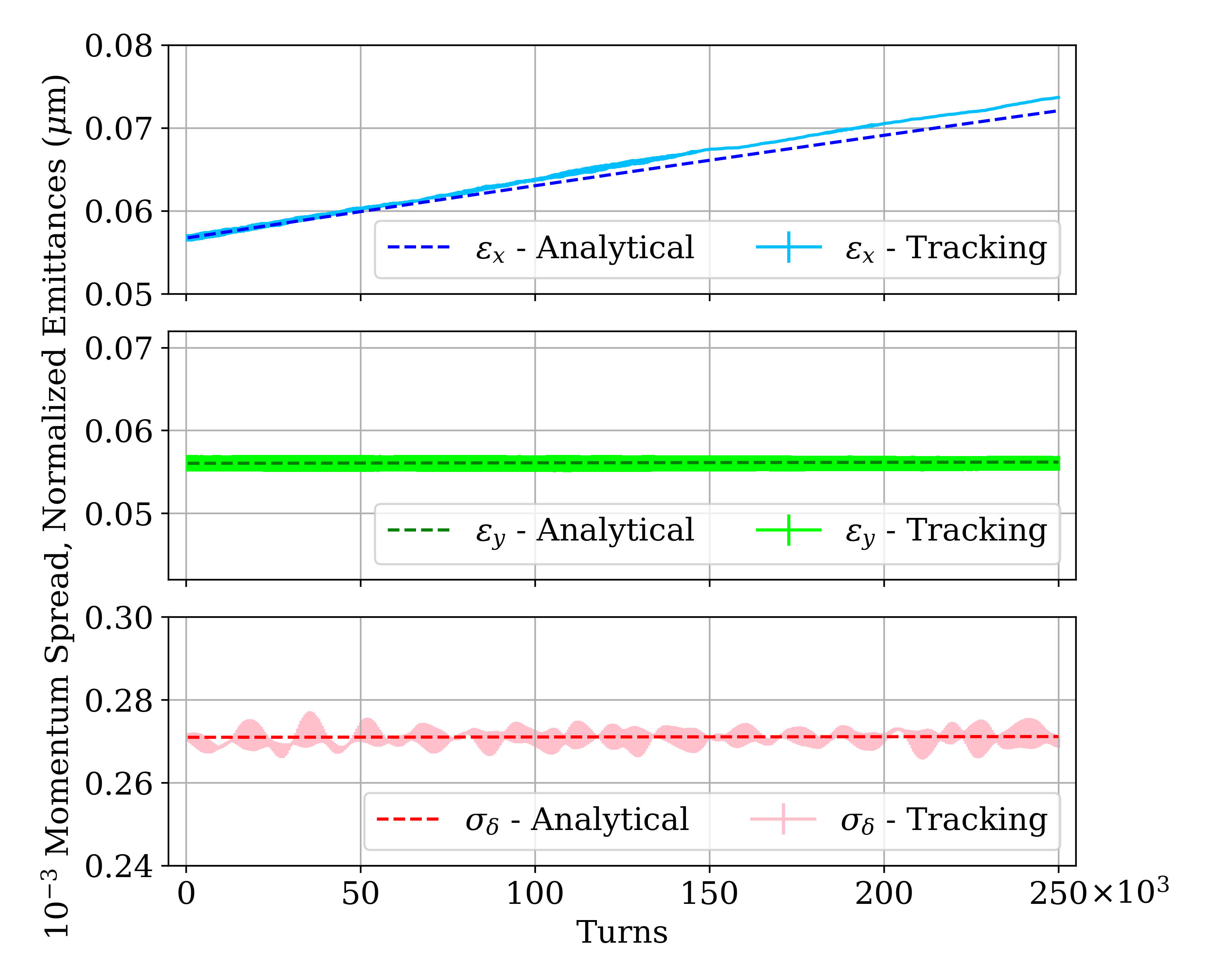}
    \caption{Comparison of the time evolution of the horizontal~(top) and vertical~(middle) emittances, and the momentum spread~(bottom), between analytical IBS predictions~(dark dashed lines) and tracking simulations including IBS~(solid lines), for LEIR operating above transition at a kinetic energy of $E_k=2.6075$~GeV/u. A round beam with normalized transverse emittances $\varepsilon_{x,y}^N\approx0.056~\mu$m is considered. Error bars indicate the standard deviation over three independent simulations.}
   \label{fig:bench_aboveT_em5}
\end{figure}

The last benchmarking case of the LEIR is the regime where the LEIR operates above transition. To accomplish that, the kinetic energy of the ions was increased to $E_k=2.6075$ GeV/u, more than 600 times the nominal value. At higher energies, the strength of the IBS effect subsides. To enhance it, in this case the transverse emittances were chosen to be very small ($\varepsilon_{x,y}^N\approx0.056~\mu$m) to result in a sizable horizontal emittance growth. Figure~\ref{fig:bench_aboveT_em5} shows the comparison between the tracking simulations with the IBS kicks and the analytic calculations, with these modifications. Excellent agreement between the simulation and the analytical prediction is observed. The relative difference between the analytical and the tracking simulations at the end is smaller than $2\%$. In this above-transition case, the qualitative behavior differs from the below-transition LEIR cases. The much higher energy strongly reduces the overall IBS strength, and the chosen parameters mainly lead to a weak horizontal emittance growth, while the vertical emittance and the momentum spread remain almost constant. The small fluctuations observed in the momentum spread are not associated with significant IBS-driven longitudinal growth, but with the much slower synchrotron motion at this energy; the synchrotron period becomes about 20000 times longer, making the synchrotron oscillations of the bunch visible in the results.

To have a reliable IBS model, it is crucial to benchmark it successfully for different ring models with different layouts and beam parameters. In this paper, two additional benchmarking cases for the SPS are presented. The ring parameters of the SPS are shown in Table~\ref{tab:sps_params}. In these SPS cases, the IBS kick is applied once per turn, while the IBS coefficients are re-evaluated every $1000$~turns according to the evolution of the transverse beam parameters and the longitudinal line density. The total duration of the simulations is $8\cdot10^5$ turns, which corresponds to 20~seconds of the actual SPS cycle.
\begin{table}[!h]
   \centering
   \caption{The SPS Parameters}
   \begin{tabular}{lc}
       \toprule
       \multicolumn{2}{c}{\textbf{Ring Parameters}}     \\
       \midrule
        Circumference [km]               & 6.9           \\
        $Pb^{82+}$ rest energy [GeV]     & 193.7         \\
        Injection Kinetic energy [GeV/u] & 5.9           \\
        Relativistic gamma, $\gamma_r$   & 7.34          \\
        Gamma transition, $\gamma_\textrm{tr}$  & 18.14         \\
        Harmonic number, $h$             & 4653          \\
        RF voltage [MV]                  & 3.2           \\
        Momentum compaction factor       & 0.003         \\
       \bottomrule
   \end{tabular}
   \label{tab:sps_params}
\end{table}

\begin{figure}[!ht]
    \centering
    \includegraphics*[trim=18 10 17 15, clip, width=1.\columnwidth]{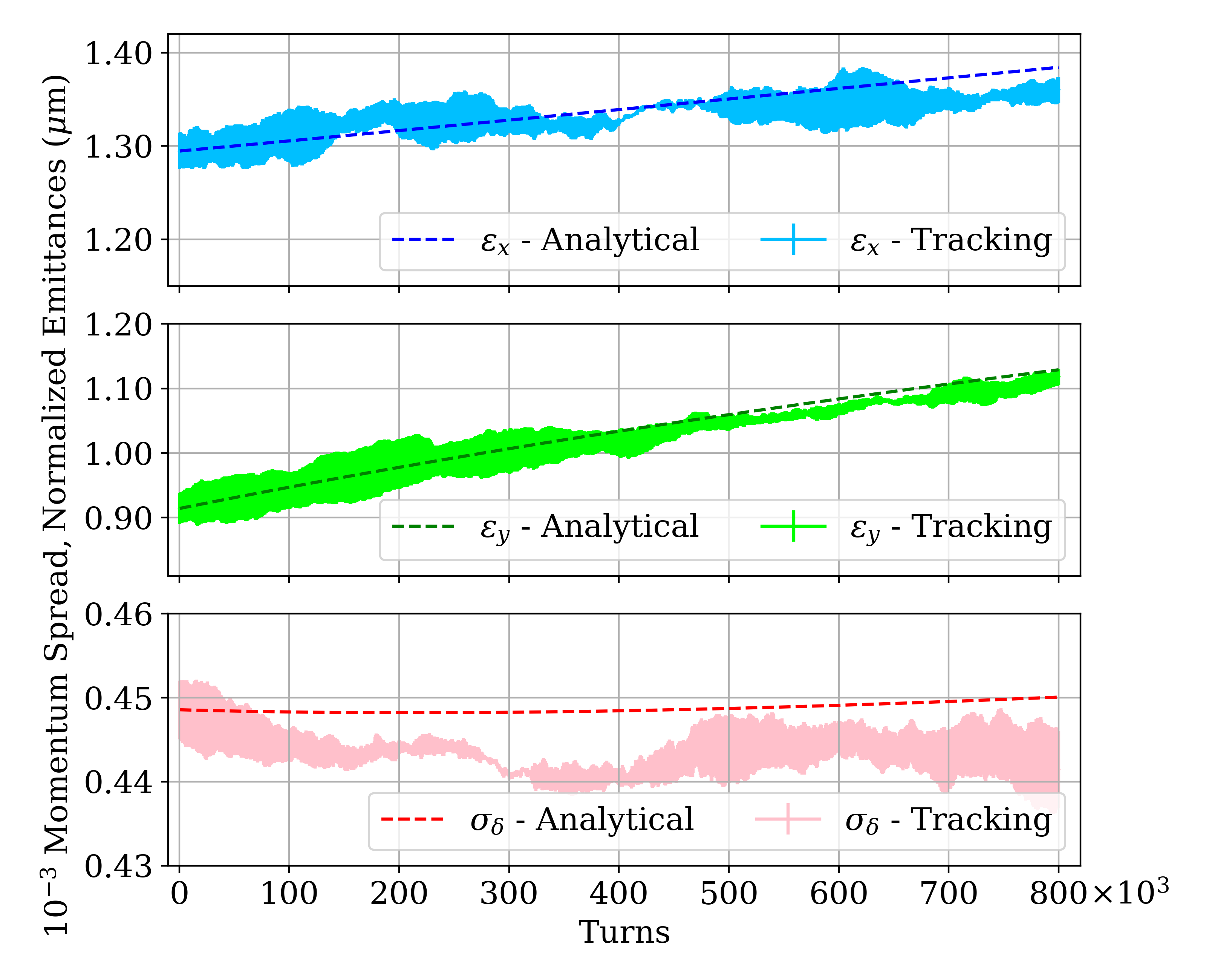}
    \caption{Comparison of the time evolution of the horizontal~(top) 
            and vertical~(middle) emittances, and the momentum spread~(bottom), between analytical IBS predictions~(dark dashed lines) and tracking simulations including IBS~(solid lines), for the SPS with initial beam parameters $\eps{x}^N=1.3~\mu$m, $\eps{y}^N=0.9~\mu$m, and a bunch length of $\sigma_z=0.23$~m. Error bars indicate the standard deviation over three independent simulations.}
   \label{fig:sps_benchmarking}
\end{figure}

The first benchmarking case is for a beam with number of charges per bunch $N_c= 2.9\times10^{10}$, normalized transverse emittances $\eps{x}^N=1.3~\mu$m, $\eps{y}^N=0.9~\mu$m and a bunch length of $\sigma_z=0.23$~m. The comparison between the tracking simulations using the modified approximate IBS model~(solid lines) and the analytical predictions based on Nagaitsev's formalism~(dark dashed lines) is shown in Fig.~\ref{fig:sps_benchmarking}. The horizontal normalized emittance is shown in the top plot, the vertical normalized emittance in the middle plot, and the momentum spread in the bottom plot. In this SPS case, both transverse emittances increase, while the momentum spread remains almost constant. The vertical emittance growth is expected because the initial vertical emittance is smaller than the horizontal one, leading to a stronger relative IBS growth in the vertical plane. At the same time, the longitudinal parameters are close to equilibrium, resulting in only a weak evolution of the momentum spread. Very good agreement is observed in all three planes, with the largest difference in the final values of about $2\%$ in the longitudinal plane.

\begin{figure}[t]
    \centering
    \includegraphics*[trim=18 10 17 15, clip, width=1.\columnwidth]{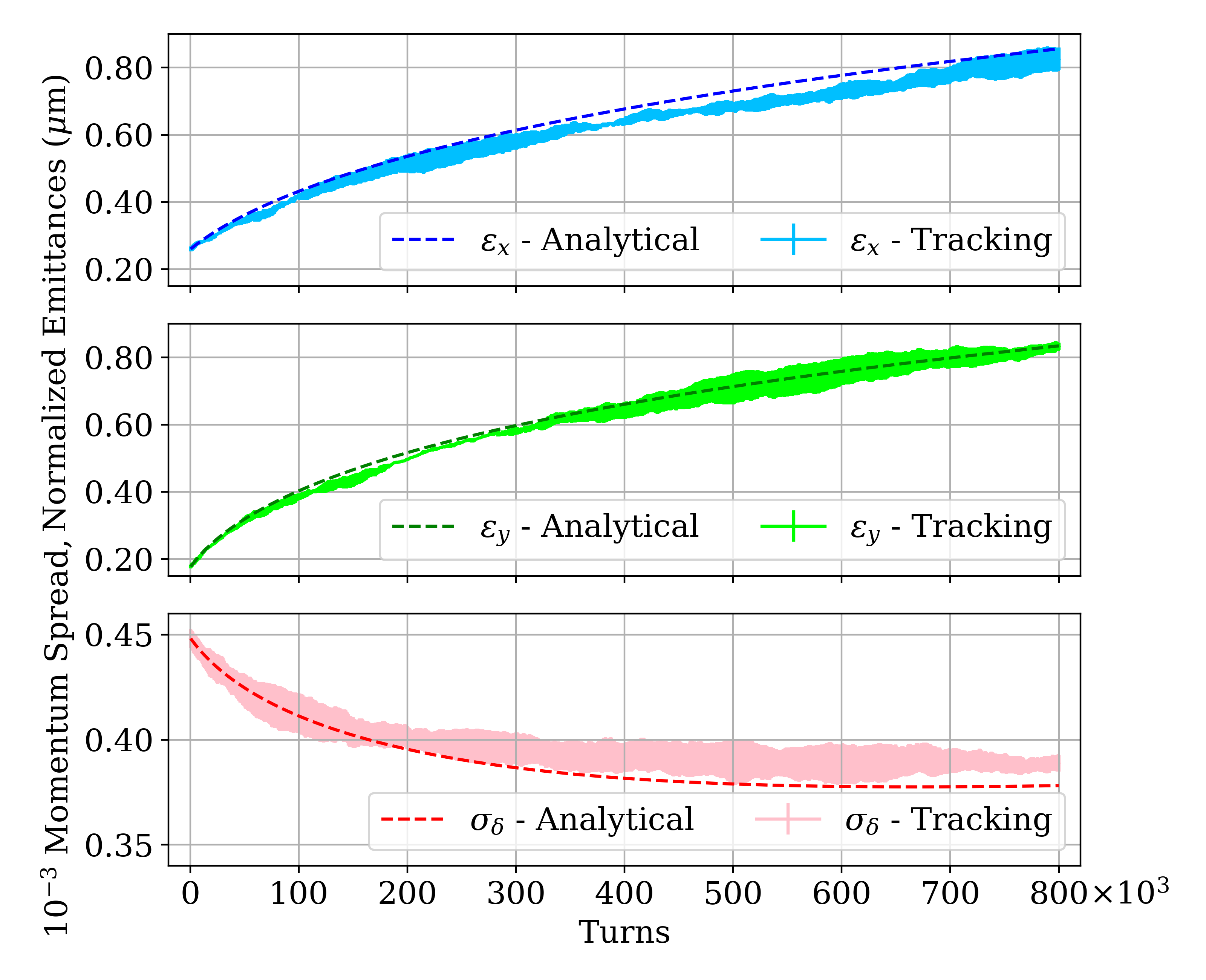}
    \caption{Comparison of the time evolution of the horizontal~(top) and vertical~(middle) emittances, and the momentum spread~(bottom), between analytical IBS predictions~(dark dashed lines) and tracking simulations including IBS~(solid lines), for the SPS with initial beam parameters $\eps{x}^N=0.26~\mu$m, $\eps{y}^N=0.18~\mu$m, and a bunch length of $\sigma_z=0.23$~m. Error bars indicate the standard deviation over three independent simulations.}
   \label{fig:sps_bench_emit_div5}
\end{figure}

To further enhance the IBS effect, a second SPS case was performed with transverse emittances one fifth of those used above. The comparison between the tracking simulations and the analytical predictions is shown in Fig.~\ref{fig:sps_bench_emit_div5}. Because of the smaller transverse emittances, the beam density is increased and the IBS effect becomes significantly stronger. In this case, both transverse emittances grow substantially, while the momentum spread decreases. This behavior is again consistent with operation below transition, where IBS can transfer part of the longitudinal momentum spread into transverse emittance growth as the beam evolves toward equilibrium. Very good agreement is observed, despite the more severe emittance exchange between the longitudinal and transverse planes. The remaining small difference is attributed to the fact that the longitudinal phase space was initialized as a binomial distribution, as in the LEIR case discussed above, while the analytical model assumes Gaussian distributions.

Six different benchmarking cases were shown, covering different regimes in LEIR and the SPS, including nominal operation below transition, stronger IBS regimes obtained by reducing the transverse emittances or the bunch length, and an above-transition LEIR case. In all cases, excellent agreement is observed between macroparticle tracking simulations employing the modified approximate IBS kick and the analytical predictions based on Nagaitsev's IBS formalism. The different qualitative behaviors of the transverse emittances and momentum spread are reproduced consistently, confirming that the stochastic kick correctly captures the IBS-driven exchange between the degrees of freedom. This level of agreement indicates that the proposed IBS model is suitable for comparisons with experimental measurements and for use in dynamical studies including additional collective effects, such as space charge. As an additional consistency check, the behavior of the approximate Piwinski invariant~\cite{Piwinski:1974it} is examined for two representative cases and compared with analytical predictions in Appendix~\ref{app:Pinv}, where the limitations of the smooth-lattice approximation are also discussed.

\section{\label{sec:5}SPACE CHARGE AND INTRABEAM SCATTERING STUDIES IN THE LEIR}
The Low Energy Ion Ring is the second accelerator and the first synchrotron of the CERN heavy-ion injector chain for the Large Hadron Collider~(LHC). The main ring parameters of the LEIR are summarized in Table~\ref{tab:LEIR_params}, as shown earlier. LEIR's magnetic cycle, as detailed in Appendix~\ref{app:LEIRcycle}, has a long injection plateau during which seven beam pulses are injected from the upstream LINAC3, at a kinetic energy of $4.2$~MeV/u. The seven pulses are stacked in the longitudinal phase space following an injection scheme which is based on a combined betatron and momentum phase space painting~\cite{LHCdesignreport}. Between the individual pulses, the stored coasting beam is compressed in the 6D phase space using electron cooling. By the end of injection, an equilibrium is established among electron cooling, IBS-driven heating, and SC-induced emittance evolution. When the electron cooler is turned off, the coasting beam is captured by the radio-frequency~(RF) cavity into two bunches. Consequently, the longitudinal line density is increased due to the bunching, which in turn leads to the enhancement of the SC and IBS effects resulting in particle loss.

To understand the mechanism for these losses and how SC and IBS influence the degradation of the beam performance of the LEIR, studies were performed for different working points in controlled conditions injecting only a single pulse from LINAC3~\cite{SaaHernandez:2019ftv}. These studies showed that the vertical emittance evolution along the duration of the magnetic cycle could not be explained by IBS only. Following up on these studies, in order to understand this discrepancy between measurements and simulations, a simulation campaign with combined IBS and SC tracking simulations was performed, to investigate the impact of the interplay between the two effects, especially in the vicinity of excited resonances, as discussed in the following.

Due to the high complexity of the machine, a set of measurements for low beam intensity (one injection), similar to~\cite{SaaHernandez:2019ftv} was chosen in order to benchmark the simulations. For these simulations, the initial parameters that were used are shown in Table~\ref{tab:beampars}. They correspond to the measured parameters right after the end of the RF-capture.
\begin{table}[!hbt]
   \centering
   \caption{Beam Parameters for the LEIR simulations}
   \begin{tabular}{lc}
        \toprule
        \multicolumn{2}{c}{\textbf{Beam Parameters}}     \\
        \midrule
        Total number of charges, $N_c$                       & 1.74$\times10^{10}$    \\
        Normalized horizontal emittance, $\varepsilon_x^N$ [$\mu$m] & 0.162    \\
        Normalized vertical emittance, $\varepsilon_y^N$ [$\mu$m]   & 0.189    \\
        Bunch length, $\sigma_z$[m]                    & 4.5     \\
       \bottomrule
   \end{tabular}
   \label{tab:beampars}
\end{table}

\begin{figure}[t]
    \centering
    \includegraphics*[trim=10 15 5 10,clip,width=1.\columnwidth]{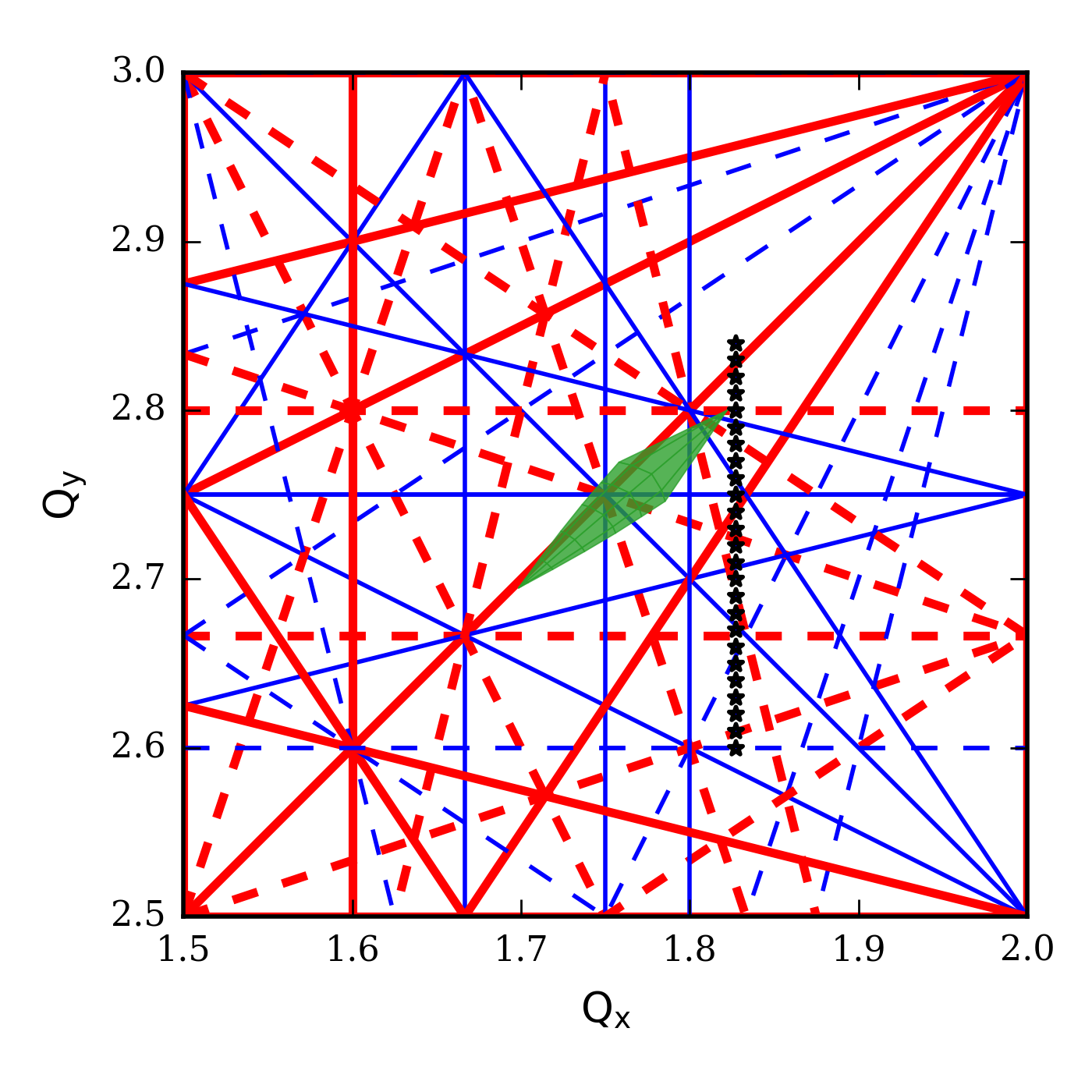}
    \caption{Illustration of the analytically calculated incoherent space-charge tune footprint for the working point $(Q_x,Q_y)=(1.82,2.80)$ in LEIR. The green-colored rhomboid represents the tune spread of on-momentum particles arising from different transverse oscillation amplitudes within the bunch, from the beam core to large amplitudes, up to $3\sigma$. The black stars indicate the vertical tune scan performed in the simulations. Normal~(solid) and skew~(dashed) systematic~(red) and nonsystematic~(blue) resonance lines up to fifth order are also shown.}
    \label{fig:footprint}
\end{figure}

A tune scan was performed in simulations using PyORBIT with the ``frozen'' potential SC kick and the ideal LEIR lattice, i.e.~without machine imperfections apart from the residual lattice perturbations remaining after the compensation of the electron cooler magnetic fields. The horizontal tune was set to $Q_x=1.82$ in all cases, while the vertical tune was varied, as indicated by the black stars in Fig.~\ref{fig:footprint}. For reference, the SC tune footprint was calculated for these beam parameters and is shown together with all the normal (full lines) and skew (dashed lines) systematic~(red) and nonsystematic~(blue) resonances up to $5^{\text{th}}$ order~\cite{Asvesta:pyscrdt}. The footprint represents the incoherent tune spread of on-momentum particles arising from different transverse oscillation amplitudes within the bunch, from the beam core to large amplitudes, and illustrates how a subset of particles can overlap with nearby resonance lines even for a fixed machine working point. The beam distribution is initialized as Gaussian in the transverse planes and as binomial in the longitudinal plane and it consists of 5000 macroparticles, which is sufficient to perform such simulations as shown by the convergence study for SPS in the Appendix~\ref{app:conv}, as well as the convergence study under extreme conditions reported in Appendix~A of~\cite{zampetakis:clic}. The aperture restrictions of the actual ring are considered~\cite{aperture:LEIR}, to accurately simulate the beam losses. Results are presented in terms of the ratios of the final to the initial values for intensity and transverse emittances. The initial values are measured 675~ms after the start of the magnetic cycle (defined as $t=0$ of the programmed cycle timing sequence), while the final ones are measured 1600~ms after the start of the magnetic cycle (for comparison, the total cycle duration including the full ramp is 3600~ms, and injection takes place at 245~ms).

\begin{figure}[h]
    \centering
    \includegraphics*[trim=22 25 20 20,clip,width=1.\columnwidth]{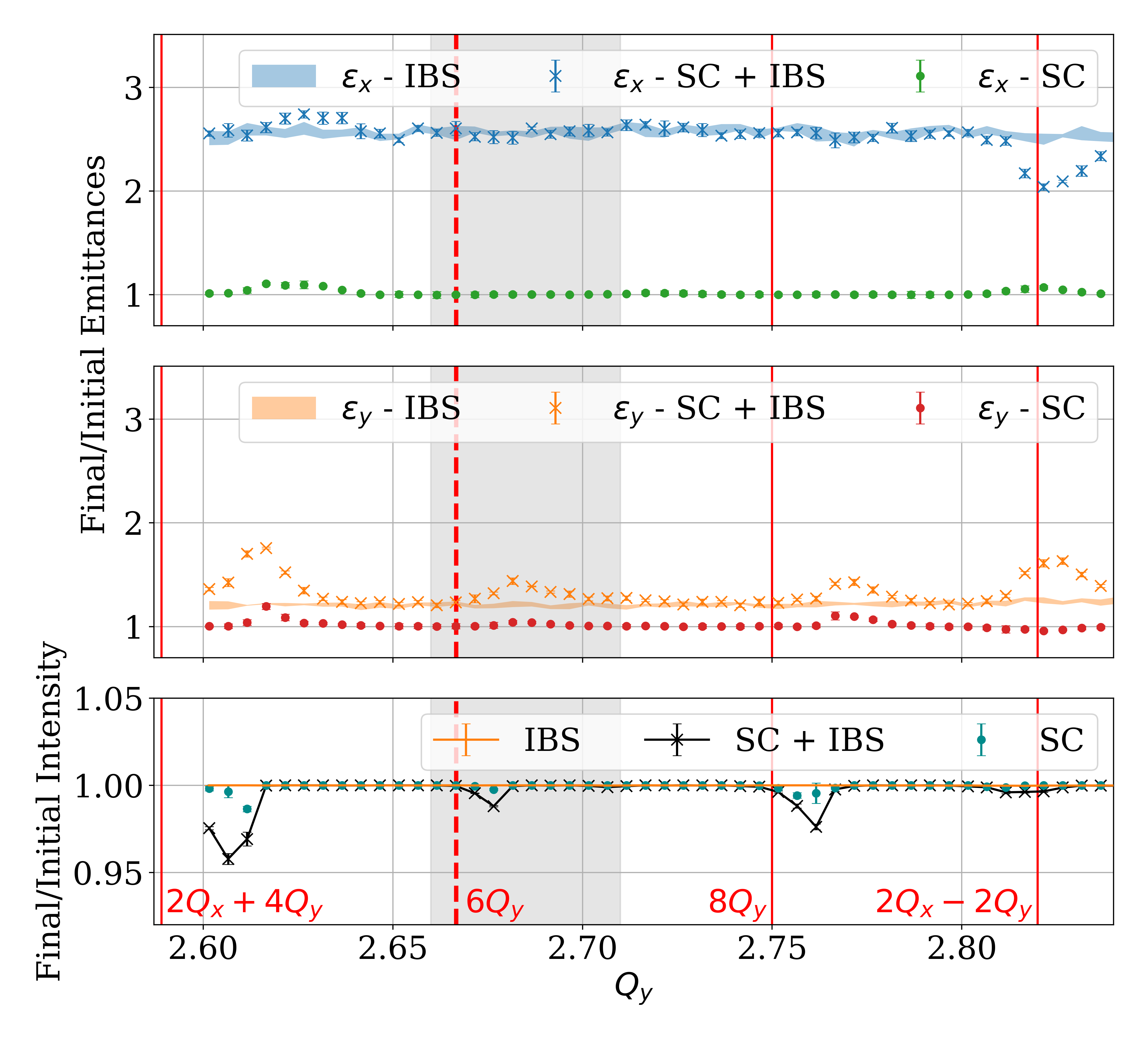}
    \caption{SC~(dots) and IBS~(lines) simulations versus combined SC and IBS simulations~(crosses) showing the final-to-initial ratios of the horizontal~(top) and vertical~(middle) emittances and intensity~(bottom) for the ideal lattice of LEIR. The error bars indicate the standard deviation over three different simulations.}
    \label{fig:sc_vs_ibs}
\end{figure}

Even in the case of this ideal lattice, the nonlinear SC potential can drive systematic resonances of even order. In the case of the LEIR, SC simulations shown with light colors in Fig.~\ref{fig:sc_vs_ibs} revealed four SC-driven resonances: the sixth order coupling resonance $2Q_x+4Q_y=14$, the sixth order vertical resonance $6Q_y=16$, the eighth order vertical resonance $8Q_y=22$ and the fourth order coupling $2Q_x-2Q_y=-2$ resonance~\cite{Asvesta:IPAC2019, PhD_Asvesta}. In these studies, we focus on the working points indicated by the grey shadowed area of the plots.

To study the interplay between SC and IBS, IBS was included in the simulations using the modified approximate IBS model described above. The results are summarized in Fig.~\ref{fig:sc_vs_ibs} and are compared to the cases where only SC and only IBS are taken into account. It is evident that in the absence of excited resonances, IBS dominates the emittance evolution, while the interplay of the two effects clearly enhances the beam response to the resonances, resulting in larger emittance blow-up and increased particle losses. 

The simulations shown in Fig.~\ref{fig:sc_vs_ibs} correspond to a purely simulation-based case study using the ideal LEIR lattice. Their purpose is to illustrate the separate and combined impact of SC and IBS, and to identify the expected beam response near the relevant resonances. For the comparison with experimental measurements, discussed below, the simulation configuration is modified to reproduce the experimental conditions more closely, including the excitation of the uncompensated $3Q_y=8$ resonance observed in LEIR.

\begin{figure}[h]
    \centering
    \includegraphics*[trim=20 25 20 20, clip, width=1.\columnwidth]{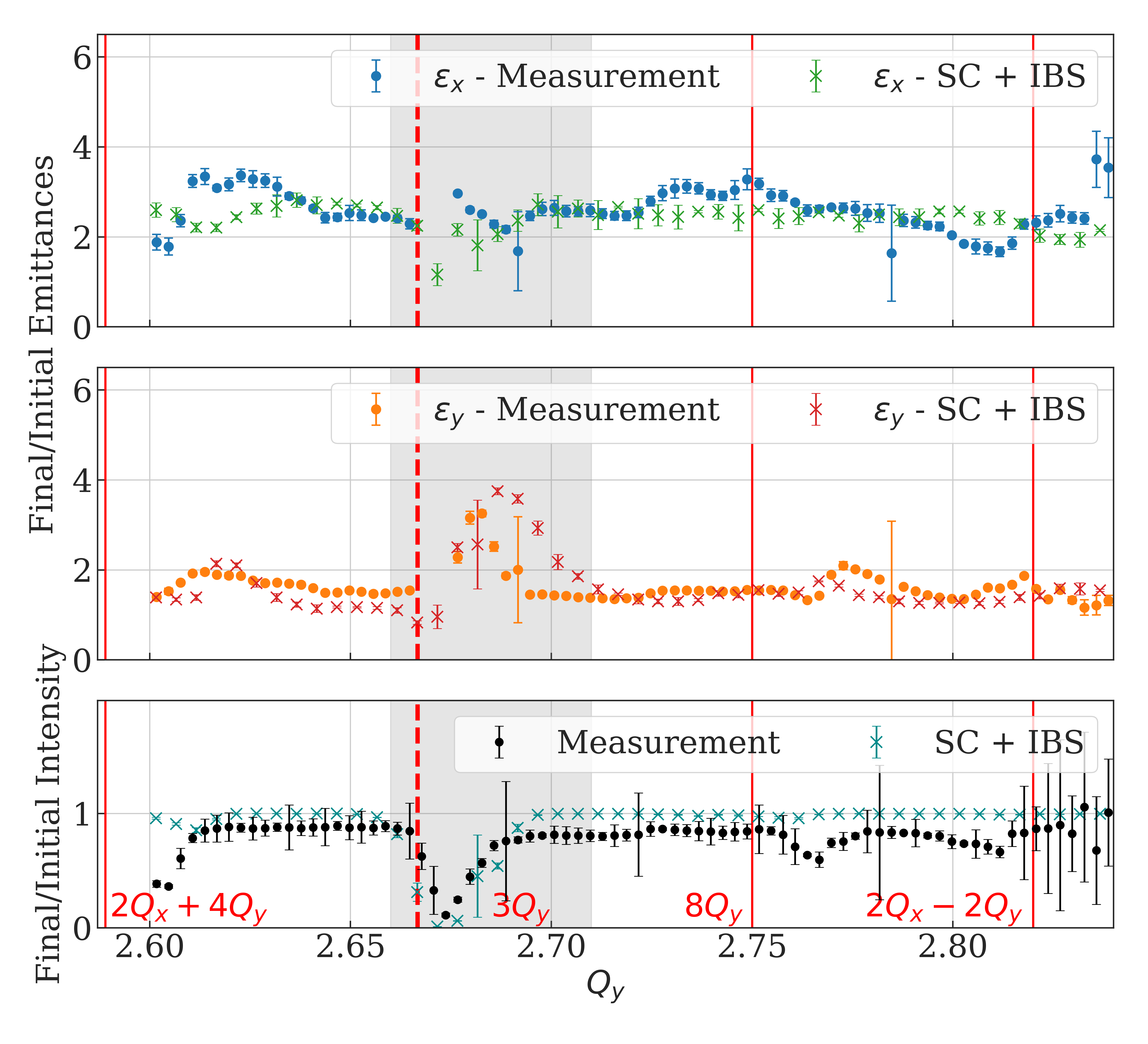}
    \caption{Measurements in LEIR~(dots) compared with combined SC and IBS simulations~(crosses), showing the final-to-initial ratios of the horizontal~(top) and vertical~(middle) emittances, and the intensity~(bottom). Error bars indicate the standard deviation over three measurements and three independent simulations, respectively.}
   \label{fig:MD_VS_IBS_SC_notcomp}
\end{figure}

To benchmark the combined SC and IBS simulations with experimental observations, a set of measurements of a vertical tune scan with a low intensity beam was used, as shown in Fig.~\ref{fig:MD_VS_IBS_SC_notcomp}. For this set of measurements the horizontal tune was set to $Q_x=1.83$ and the results are presented in terms of the ratios of the final to the initial values of the transverse emittances and the intensity. As before, the initial values are measured 675~ms after the start of the magnetic cycle, i.e.~the end of the RF-capture, while the final ones are measured 1600~ms after the start of the magnetic cycle. For each measurement a fresh beam was injected.

\begin{figure}[h]
    \centering
    \includegraphics*[trim=25 15 18 18,clip,width=1.\columnwidth]{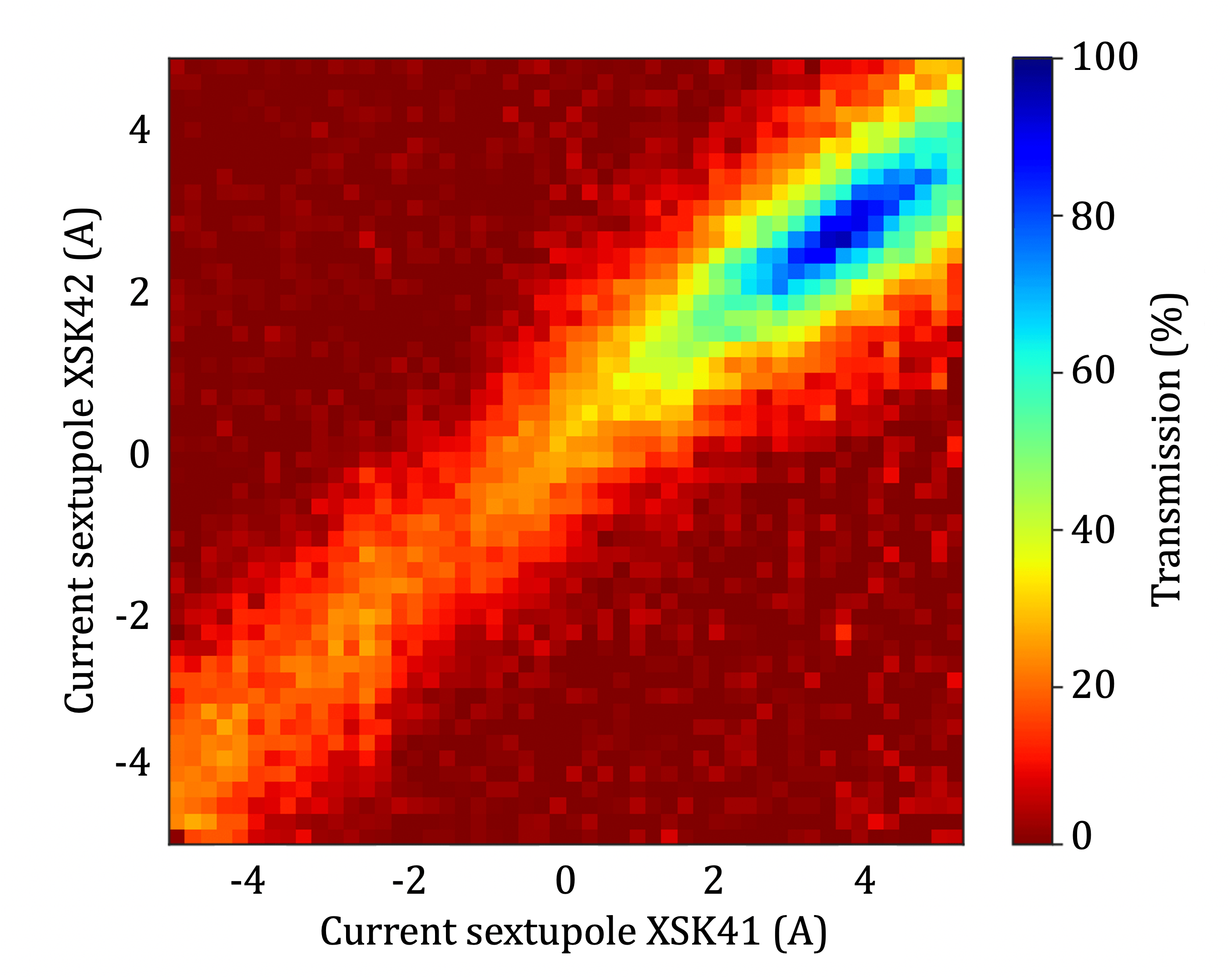}
    \caption{Beam transmission in LEIR while crossing the $3Q_y=8$ resonance as a function of the currents in the skew sextupoles~\cite{SaaHernandez:2019ftv}.}
   \label{fig:sextset}
\end{figure}

Due to the low lattice periodicity (2-fold) of the LEIR and due to random magnetic errors, various resonances are observed to be excited in the measurements, including the third order vertical $3Q_y=8$ and the third order coupling $Q_x+2Q_y=7$ resonances. Attempts have been made to compensate the $3Q_y=8$ resonance by using two independently powered skew sextupole correctors with the appropriate phase advance~\cite{SaaHernandez:2019ftv}. The compensation was optimized by maximizing the beam transmission while crossing the excited resonance dynamically. The results are shown in Fig.~\ref{fig:sextset} in terms of the current of the skew sextupoles and the resulting beam transmission during the resonance crossing. Corrector settings could be found that restore a beam transmission of more than 90\% compared to 30\% without compensation.

To re-create the uncompensated resonance of the experiment in the simulation study, skew sextupole errors were added in the ideal lattice to excite the $3Q_y$ resonance. A suitable error for this case is to use the optimal settings of the skew sextupole correctors that compensate the $3Q_y$ resonance in the real machine, but with inverted polarity. This error primarily affects the $3Q_y$ resonance and is therefore expected to produce a similar resonance response in the simulations and measurements.

Combined simulations of IBS and SC were performed using this configuration for a horizontal tune of $Q_x=1.83$ and a simulation duration matching the $925$~ms measurement window. For improved readability, only the results from simulations including both SC and IBS are shown here; the corresponding cases with only SC or only IBS, performed under the same conditions, are presented in Appendix~\ref{app:LEIRsim} for completeness. The macroparticle distribution was again initialized as Gaussian in the transverse planes and binomial in the longitudinal plane, consisting of 5000 macroparticles. The results are presented in Fig.~\ref{fig:MD_VS_IBS_SC_notcomp} in comparison with the measurements. The simulations reproduce the measured intensity minimum near the $3Q_y=8$ resonance and the main qualitative features of the transverse-emittance response, although quantitative differences remain in the emittance ratios. The intensity minimum does not coincide exactly with the nominal resonance condition because the SC-induced tune spread causes part of the beam to encounter the resonance when the machine working point lies slightly above it. However, in the real machine there seem to be errors exciting additional resonances in the measurements that do not clearly appear in the simulation. Thus, these resonances seem not to be caused by the space charge potential itself, but rather from additional unknown magnetic errors present in the ring, but not reflected in the corresponding model.

\section{\label{sec:6}SPACE CHARGE AND INTRABEAM SCATTERING STUDIES FOR THE SPS}
The SPS is the largest accelerator of the LHC ion injector chain, with a circumference of around 7~km. A more detailed list of parameters was shown earlier in Table~\ref{tab:sps_params}. Despite the thousandfold higher kinetic energy of the Pb ion beam at the SPS injection compared to the LEIR injection, SC induces a considerable tune shift of $\Delta Q_{x,y}=(-0.2,-0.29)$ due to the short bunch length and the large machine circumference, making the beam susceptible to resonances.

\begin{figure}[h]
    \centering
    \includegraphics*[trim=17 5 10 5,clip,width=1.\columnwidth]{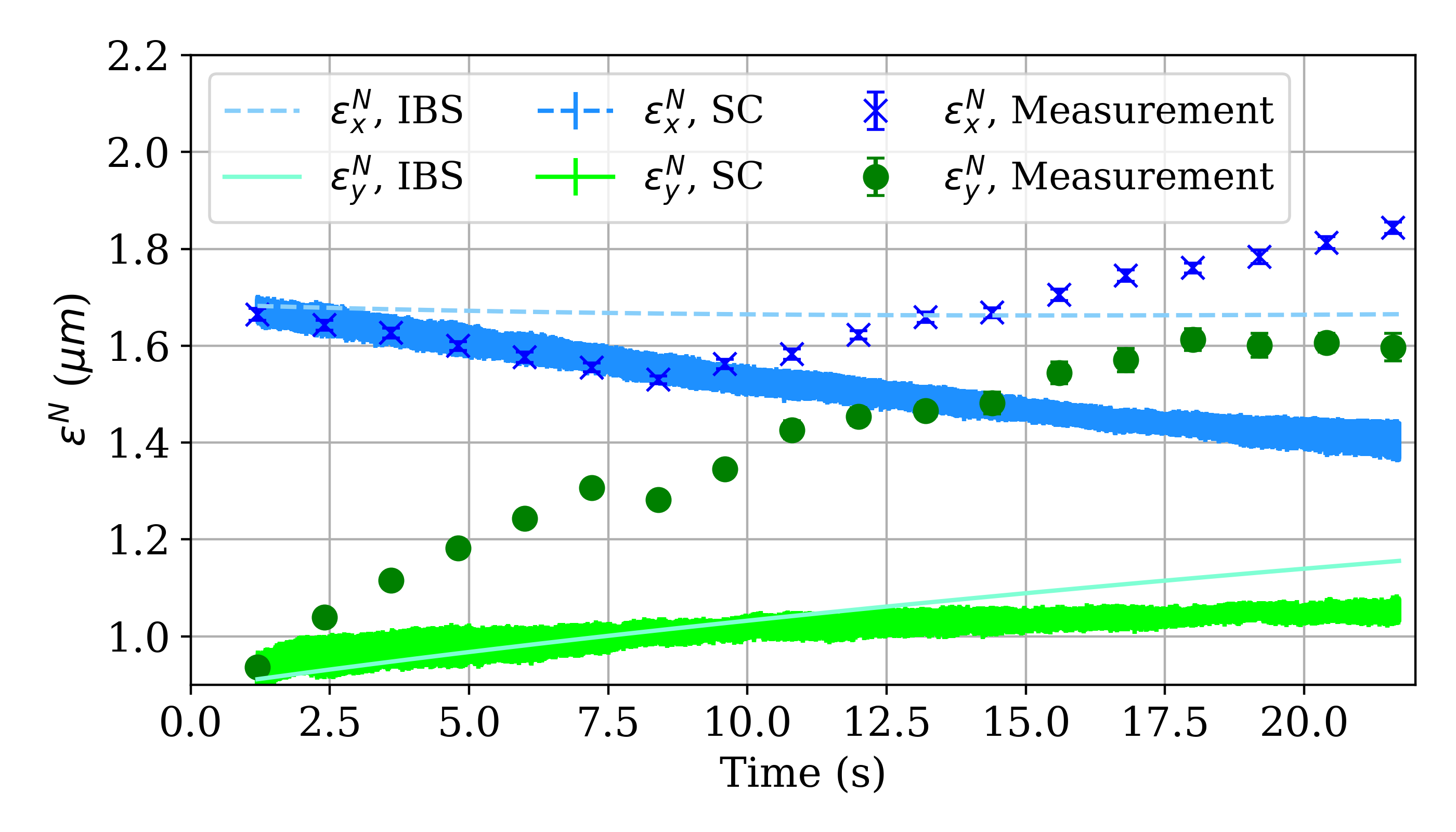}
    \caption{Evolution of the horizontal~(blue) and vertical~(green) emittance as observed in operation~(markers), from analytical IBS calculations using Nagaitsev's method (light-colored lines) and from SC simulations (lines with error bars) in PyORBIT (averaged over three independent runs). The error bars in the simulations correspond to one standard deviation, while those in the measurements correspond to the profile fitting error.}
    \label{fig:SPS_indi}
\end{figure}

In the operational conditions of 2016, an emittance exchange between horizontal and vertical planes, followed by a large emittance blow-up in both planes, was observed along the flat bottom of the SPS acceleration cycle, as shown in Fig.~\ref{fig:SPS_indi}~(markers). A campaign of standalone IBS calculations and SC simulations was performed in 2020~\cite{IJohn} as an attempt to explain the observed emittance evolution. The SC-only and combined SC and IBS tracking simulations used the same general configuration as for LEIR, with the difference that the SC potential was re-evaluated every 1000~turns based on the evolution of the tracked particles. In the combined simulations, the IBS coefficients were also re-evaluated every 1000~turns, while the IBS kick was applied every turn to each particle, according to the beam parameters and the particle's longitudinal position.

The generated initial distribution is Gaussian in all planes and comprises 1000 macroparticles, which is a necessary compromise given the high complexity of these simulations, but is sufficient as demonstrated by the convergence studies in Appendix~\ref{app:conv}. In addition, a quadrupolar error is included in the lattice model, inducing a 5--10\% beta-beating, similar to what is observed in the real machine under operational conditions. To accurately simulate the beam losses, the realistic physical aperture model of the SPS is taken into account in the simulations~\cite{aperture:SPS}.

The results of the SC simulation~(dark lines with error bars) and the analytical IBS predictions using Nagaitsev's method~(light lines) are also shown in Fig.~\ref{fig:SPS_indi}. It is evident that the observed emittance behavior cannot be explained by either standalone IBS calculations or SC simulations.

\begin{figure}[h]
    \centering
    \includegraphics*[trim=18 5 10 5,clip,width=1.\columnwidth]{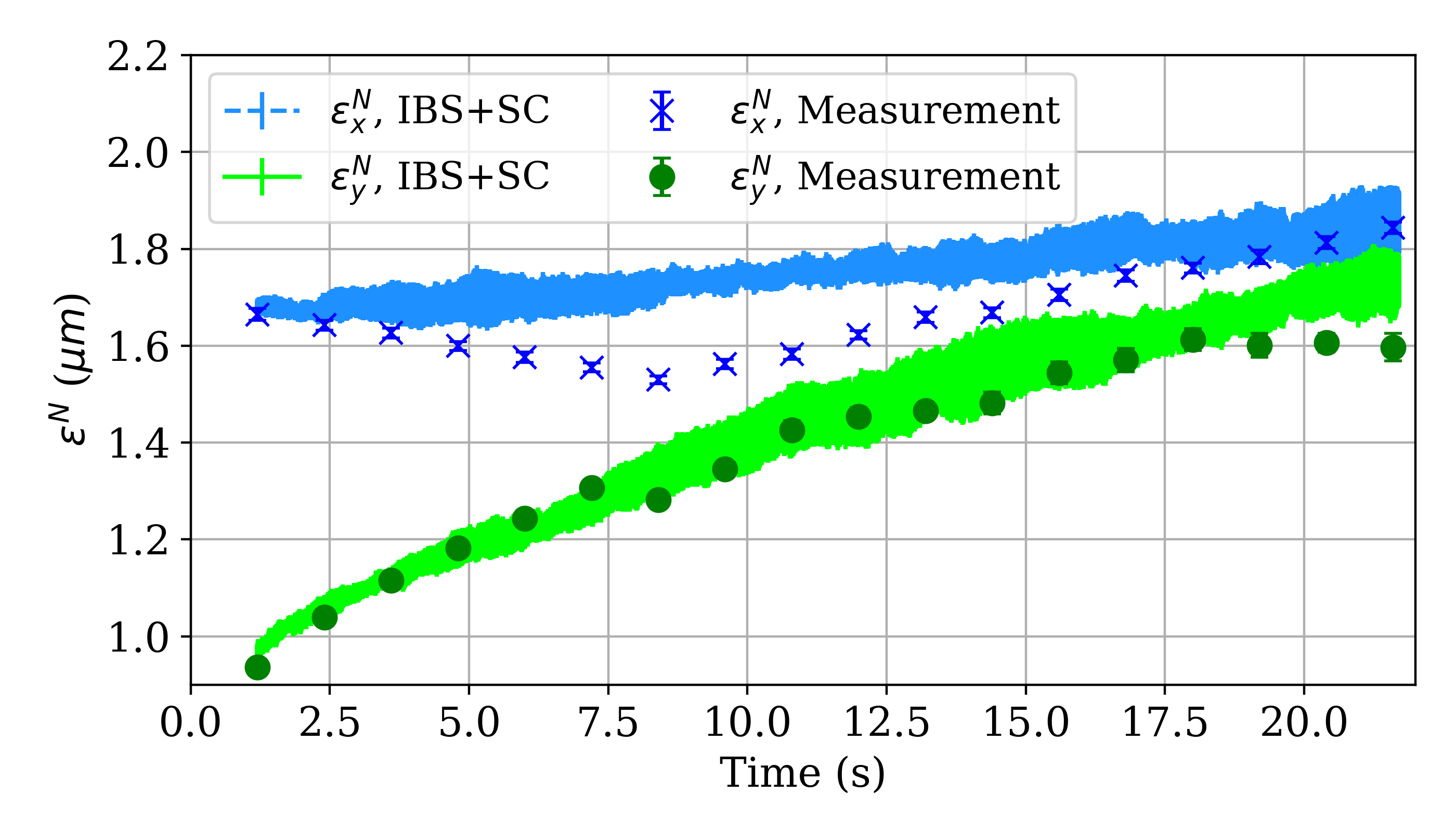}
    \caption{Evolution of the horizontal~(blue) and vertical~(green) emittance as observed in operation~(markers) and from combined SC and IBS simulations~(lines) in PyORBIT (averaged over three independent runs). The error bars in the simulations correspond to one standard deviation, while those in the measurements correspond to the profile fitting error.}
   \label{fig:SPS_kinetic}
\end{figure}

To this end, using the aforementioned setup, a simulation campaign was initiated to investigate the impact of the interplay between SC and IBS and whether this interplay could explain the observed emittance behavior. Figure~\ref{fig:SPS_kinetic} shows the evolution of the normalized horizontal~(blue) and vertical~(green) emittances in time, as observed from the measurements~(markers) and as predicted by the combined simulations~(solid lines). This is the first simulation study to reproduce the measured emittance evolution in both transverse planes with good overall agreement, demonstrating the importance of the interplay between the two effects in this case. The agreement is particularly good in the vertical plane, while a minor discrepancy remains in the horizontal trend. It should be noted that the experimentally observed particle losses were of the order of 10\% and they were not recreated in these simulations, which might explain the slightly larger emittance growth. Although no data were stored for the evolution of the longitudinal beam profile during these measurements, there is typically a reduction in bunch length observed experimentally. In contrast, in the simulations presented here there is instead an increase of the bunch length by about 6\%. These differences need to be addressed in more detail in future studies.

\section{\label{sec:7}Discussion on the interplay between space charge and intrabeam scattering}

A detailed analysis of the microscopic mechanism behind the observed interplay between SC and IBS is beyond the scope of this work. However, the simulations presented here already provide some qualitative insight into its nature.

The SC force introduces a significant incoherent tune spread within the bunch, which can bring particles close to resonance conditions. As a result, SC-driven periodic resonance crossing enhances transverse diffusion, leading to emittance growth and particle losses, particularly near low-order resonances. IBS, on the other hand, acts as a stochastic diffusion mechanism, continuously exchanging energy among particles through multiple small-angle Coulomb collisions. When both effects act simultaneously, the IBS-induced diffusion can enhance the SC-driven resonance crossing, allowing particles to reach larger amplitudes more rapidly and amplifying the emittance growth and loss rates. For instance, on-momentum particles that would otherwise not satisfy the resonance condition may intermittently cross it due to the perturbations introduced by IBS.

No analytical model is presently available for the coupled nonlinear dynamics of SC, IBS, lattice resonances, and particle losses. Therefore, this interplay is studied here using macroparticle tracking simulations. The present study represents a first step in identifying and characterizing this interplay. A more detailed investigation of the underlying mechanisms and their parametric dependencies will be the subject of future work.

\section{\label{sec:8}CONCLUSIONS AND FUTURE PLANS}
In this paper, an IBS kick was implemented in a tracking code to study the interplay between SC and IBS in the context of the LEIR and SPS accelerators for Pb ion beams. Both accelerators operate below transition, which makes the implementation of IBS in tracking simulations not trivial. 

The existing IBS approximate model based on the kinetic theory of gases, implemented by Zenkevich, Bolshakov and Boine-Frankenheim, is capable of simulating the IBS effect in macroparticle tracking simulations using a significantly smaller number of macroparticles compared to the BCM model, which requires a large number of macroparticles to sufficiently populate all the cells. Inspired by this model, a more general extension is presented in this publication that is able to simulate the IBS effect for a wider range of accelerators, using the formalism of Nagaitsev and including the longitudinal line density to modulate the kicks for the particles according to their position within the bunch. Benchmarking cases for the LEIR and the SPS are presented, for different parameter regimes. All benchmarking cases show excellent agreement with the analytical predictions of IBS. 

For the case of the LEIR, standalone SC and IBS simulations as well as simulations combining both effects were performed using the ideal lattice (i.e.~without machine imperfections). In the LEIR, several SC-driven resonances are excited due to the low lattice periodicity and the nonlinear potential from SC for a Gaussian transverse distribution. The combined simulations show a clear enhancement of the particles' response to the resonances, leading to larger emittance blow-up and increased particle losses. An attempt was made to compare the combined simulations with experimental data by adding an error to the ideal lattice in the simulations to excite the $3Q_y$ resonance, equivalent to the residual nonlinear error remaining after the correction applied in the real machine to compensate it. 

The comparison between these simulations and the measurements in which the $3Q_y=8$ resonance is uncompensated shows that the interplay of IBS and SC reproduces well the measured intensity losses and captures the main features of the emittance evolution in both transverse planes for the tunes most affected by the resonance components. The quantitative differences in the emittance values may reflect both the limited precision of the transverse-profile measurements and limitations of the current simulation model, including lattice imperfections that are not fully represented. In addition, particle losses appear to be generally larger for all working points in the real machine.

For the SPS, standalone SC simulations and IBS calculations, as well as combined SC and IBS simulations, were performed in the presence of a quadrupolar error inducing a 5--10\% beta-beating, similar to what is observed in the real ring under operational conditions. Neither the standalone SC simulations nor the IBS calculations could explain the beam behavior observed in the 2016 measurements. By contrast, the combined SC and IBS simulations reproduced the main features of the measured transverse-emittance evolution from 2016, highlighting the importance of the interplay between the two effects. 

Our study shows that the interplay of SC and IBS effects needs to be taken into account in order to reproduce experimental observations of beam degradation close to lattice resonances for both the LEIR and the SPS. In addition, a good knowledge of the lattice imperfections in both rings is important for obtaining an accurate performance prediction and optimization.

\section*{ACKNOWLEDGMENTS}
The authors would like to thank N.~Biancacci and the other members of the LEIR team at CERN for the insight and the support given during the machine experiments, and A.~Latina for fruitful discussions.

\appendix

\section{Bjorken--Mtingwa formalism and kinetic derivation of the IBS kick}
\label{app:BM_kinetic}
For completeness, this appendix summarizes the Bjorken--Mtingwa formalism and the original approximate kinetic model underlying the stochastic IBS kick. These expressions provide the theoretical background for the modified kick introduced in the main text. The simulations presented in this work use the computationally efficient implementation based on Nagaitsev's formalism, while the corresponding construction of the effective kick coefficients using the Bjorken--Mtingwa formalism is given here for reference.

\subsection{Bjorken--Mtingwa formalism}
\label{app:BM}
Following the classical IBS model of Piwinski~\cite{Piwinski:1974it}, Bjorken and Mtingwa~(BM) introduced a formalism to analytically estimate the effect of IBS in beams of charged particles~\cite{Bjorken:0}. Their model is based on Fermi's relativistic ``Golden Rule'' for the transition rate due to a two-body scattering process and includes the effect of strong-focusing lattices.

In the BM model, the emittance growth rates due to IBS are characterized by the auxiliary matrices $L^{(x)}$, $L^{(y)}$, $L^{(z)}$. After some years, these matrices have been generalized to include non-relativistic factors and vertical dispersion~\cite{Conte:1985, Antoniou:mad}. The rows and columns of these matrices follow the ordering $(x,z,y)$, where $z$ denotes the longitudinal direction. The updated matrices are given by:
\begin{equation}
    L^{(x)}=
    \begin{pmatrix}
        \cfrac{\beta_x}{\eps{x}} & -\gamma\F{x}\cfrac{\beta_x}{\eps{x}} & 0~  \\
        -\gamma\F{x}\cfrac{\beta_x}{\eps{x}}~ & \gamma^2\cfrac{\cH{x}}{\eps{x}} ~& 
        0~ \\
        0 & 0 & 0~  \\
    \end{pmatrix},
\end{equation}
\begin{equation}
    L^{(y)}=
    \begin{pmatrix}
        ~0 & 0 & 0 \\
        ~0 & \gamma^2\cfrac{\cH{y}}{\eps{y}} ~& -\gamma\F{y}\cfrac{\beta_y}{\eps{y}} \\
        ~0 & -\gamma\F{y}\cfrac{\beta_y}{\eps{y}} & \cfrac{\beta_y}{\eps{y}} \\
    \end{pmatrix},
\end{equation}
\begin{equation}
    L^{(z)}=
    \begin{pmatrix}
        0 & 0 & 0 \\
        0 & \cfrac{\gamma^2}{\sigma_{\delta}^2} & 0 \\
        0 & 0 & 0 \\
    \end{pmatrix}.
\end{equation}

The sum of these three matrices gives the general auxiliary matrix $L$, defined as:
\begin{equation}\label{eq:L}
    L=
    \begin{pmatrix}
        \cfrac{\beta_x}{\eps{x}} & -\gamma\F{x}\cfrac{\beta_x}{\eps{x}} & 0 \\
        -\gamma\F{x}\cfrac{\beta_x}{\eps{x}}~~~ & \gamma^2(\cfrac{\cH{x}}{\eps{x}}+\cfrac{\cH{y}}{\eps{y}}+\cfrac{1}{\sigma_{\delta}^2}) ~~~& -\gamma\F{y}\cfrac{\beta_y}{\eps{y}} \\
        0 & -\gamma\F{y}\cfrac{\beta_y}{\eps{y}} & \cfrac{\beta_y}{\eps{y}}  \\
    \end{pmatrix},
\end{equation}
where $\gamma$ is the Lorentz factor, $\eps{x,y}$ are the transverse emittances, $\sigma_{\delta}$ is the momentum spread, $\beta_{x,y}$ are the beta functions of the lattice, $\eta_{x,y}$ and $\eta'_{x,y}$ are the dispersion functions and their derivatives, and $\F{x,y}$ and $\cH{x,y}$ are functions of the optics parameters given by
\begin{equation}
    \F{x,y}=\eta'_{x,y}-\frac{\beta'_{x,y}\eta_{x,y}}{2\beta_{x,y}},
\end{equation}
\begin{equation}
    \cH{x,y}=\frac{1}{\beta_{x,y}}\Bigg[\eta^2_{x,y}+\Bigg(\beta_{x,y}\eta'_{x,y}-
    \frac{1}{2}\beta_{x,y}'\eta_{x,y}\Bigg)^2\Bigg].
\end{equation}

The growth rates can be expressed through the diffusion kernels $I_{ij}^\mathrm{BM}$ ($K_{ij}$ in~\cite{Bjorken:0}), defined as:
\begin{equation}\label{eq:I_theta}
    I_{ij}^{\mathrm{BM}}
    =
    \frac{A_\mathrm{BM}}{4\pi}
    \int
    \frac{
        d^3\theta\,
        \exp\!\left(
            -\frac{1}{4}
            \boldsymbol{\theta}^{\mathsf T}
            L
            \boldsymbol{\theta}
        \right)
    }{\theta^3}
    \left(
        \delta_{ij}\theta^2
        -3\theta_i\theta_j
    \right),
\end{equation}
where $\boldsymbol{\theta}=(\theta_x,\theta_z,\theta_y)^{\mathsf T}$, $\theta^2=\boldsymbol{\theta}^{\mathsf T}\boldsymbol{\theta}$, and $\theta_i$ is the scattering angle in plane $i$. The constant $A_\mathrm{BM}$ is defined as
\begin{equation}\label{eq:A_BM}
    A_\mathrm{BM} = \frac{cNr_0^2}{8\pi\beta^3\gamma^4\eps{x}\eps{y}\sigma_z\sigma_\delta}  \clog,
\end{equation}
where $\clog$ is the Coulomb logarithm, $\beta$ and $\gamma$ are the relativistic parameters, $c$ is the speed of light in vacuum and $r_0$ is the classical radius of the particle.

To make the integral expressions simpler, the kernels can be expressed explicitly in terms of the tensor $\mathscr{L}=L+\lambda I$, with $I$ the unit matrix, following~\cite{Bjorken:0}:
\begin{equation}\label{eq:I_BM}
    I_{ij}^\mathrm{BM} = A_\mathrm{BM} \int_{0}^{\infty} \frac{d\lambda~\lambda^{1/2}}{\sqrt{\det\mathscr{L}}}
    (\delta_{ij}\mathrm{Tr}\mathscr{L}^{-1}-3\mathscr{L}^{-1}_{i,j}).
\end{equation}

Then, the growth rates are obtained by summing the diffusion kernels with the auxiliary matrices $L^{(i)}$, with $i=x,y,z$:
\begin{equation}
    \frac{1}{T_{x,y,z}} = \left\langle \sum_{i,j}I_{ij}^{\mathrm{BM}}L_{ij}^{(x,y,z)} \right\rangle,
\end{equation}
which, if we extend the summation, leads to the following expressions for the growth rates in each plane:
\begin{equation}
    \frac{1}{T_x} = \Big\langle\frac{\beta_x}{\eps{x}}I_{xx}^\mathrm{BM}-2\gamma
    \frac{\beta_x\F{x}}{\eps{x}}I_{xz}^\mathrm{BM}+\gamma^2\frac{\cH{x}}{\eps{x}}I_{zz}^\mathrm{BM}
    \Big\rangle,
    \label{eq:Tx}
\end{equation}
\begin{equation}
    \frac{1}{T_y} = \Big\langle\frac{\beta_y}{\eps{y}}I_{yy}^\mathrm{BM}-2\gamma
    \frac{\beta_y\F{y}}{\eps{y}}I_{yz}^\mathrm{BM}+\gamma^2\frac{\cH{y}}{\eps{y}}I_{zz}^\mathrm{BM}
    \Big\rangle,
    \label{eq:Ty}
\end{equation}
\begin{equation}
    \frac{1}{T_z} = \Big\langle\frac{\gamma^2}{\sigma_{\delta}^2}I_{zz}^\mathrm{BM}\Big\rangle,
    \label{eq:Tp}
\end{equation}
where the brackets denote averaging around the ring circumference.

\subsection{Original approximate kinetic model}
\label{app:kinetic}
The kinetic starting point and the corresponding Langevin equation are introduced in Sec.~\ref{sec:3}. Here, we recall the friction and diffusion coefficients used in the original approximate model of Refs.~\cite{Zenkevich:0,Zenkevich:2}. These coefficients are obtained by averaging over the particle distribution and assuming that most IBS interactions occur in the beam core. The friction force is taken to be linear in momentum, and the diffusion tensor is assumed to be constant. Under these assumptions, the friction and diffusion coefficients are written as
\begin{equation}\label{eq:K_theta}
    K_i=\frac{A_\mathrm{BM}}{\left< p_i^2\right>}\mathcal{I}_{i,i},
\end{equation}
\begin{equation}\label{eq:D_theta}
    D_{i,j} = A_\mathrm{BM} \left( \delta_{ij}\sum_{k=1}^3 \mathcal{I}_{k,k} - \mathcal{I}_{i,j} \right).
\end{equation}
In this model, it is assumed that
\begin{equation}
    \left< p_{x}^2\right> = \frac{\eps{x}}{\beta_x}, \qquad
    \left< p_{y}^2\right> = \frac{\eps{y}}{\beta_y}, \qquad
    \left< p_{z}^2\right> = \frac{\sigma_{\delta}^2}{\gamma^2}.
\end{equation}
The integrals $\mathcal{I}_{i,j}$ are given by
\begin{equation}
    \mathcal{I}_{i,j}
    =
    \int
    \frac{
        d^3\theta\,
        \exp\!\left(
            -\frac{1}{4}
            \boldsymbol{\theta}^{\mathsf T}
            L
            \boldsymbol{\theta}
        \right)
    }{\theta^3}
    \theta_i\theta_j .
\end{equation}
The relation between the diffusion coefficients and the noise-amplitude matrix $C_{i,j}$ entering the Langevin equation follows the procedure described in Ref.~\cite{Zenkevich:0}.

\subsection{Effective kick coefficients using the Bjorken--Mtingwa formalism}
\label{app:BM_modified}

Using the friction and diffusion coefficients defined in Eqs.~\eqref{eq:K1} and~\eqref{eq:D1}, together with the relation between the BM diffusion kernels and the kinetic coefficients given in Eq.~\eqref{eq:IKD}, the analytical IBS growth-rate expressions of Appendix~\ref{app:BM} can be rewritten in terms of effective diffusion and friction coefficients. This gives the BM-based counterpart of the coefficients used in the modified stochastic kick of Eq.~\eqref{eq:mod_LE}. Since the friction matrix is assumed to be diagonal in the approximate kinetic model, only the diagonal friction coefficients \(K_i\) are retained in the expressions below.

{After substituting Eq.~\eqref{eq:IKD} into Eqs.~\eqref{eq:Tx}--\eqref{eq:Tp}, each growth rate can be written as the difference between a diffusion contribution and a friction contribution,
\begin{equation}
    \frac{1}{T_u}=G_u-F_u,\qquad u=x,y,z.
\end{equation}
The terms proportional to the diffusion coefficients \(D_{i,j}\) define \(G_u\), while the terms proportional to the diagonal friction coefficients \(K_i\) define \(F_u\). This separation gives the effective coefficients used in the modified Langevin kick.
\begin{equation}\label{bm:Gx}
    G_x = \Big\langle\frac{\beta_x}{\varepsilon_x}D_{x,x}-6\gamma\frac{\beta_x\F{x}}{\varepsilon_x}D_{x,z}+\gamma^2\frac{\cH{x}}{\varepsilon_x}D_{z,z}\Big\rangle,
\end{equation}
\begin{equation}
    G_y = \Big\langle\frac{\beta_y}{\varepsilon_y}D_{y,y}-6\gamma\frac{\beta_y\F{y}}{\varepsilon_y}D_{y,z}+\gamma^2\frac{\cH{y}}{\varepsilon_y}D_{z,z}\Big\rangle,
\end{equation}
\begin{equation}
    G_z = \Big\langle\frac{\gamma^2}{\sigma_{\delta}^2}D_{z,z}\Big\rangle,
\end{equation}
\begin{equation}
    F_x = \Big\langle
    2K_x
    +2\gamma^2\sigma_{\delta}^2
    \frac{\cH{x}}{\varepsilon_x}K_z
    \Big\rangle,
\end{equation}
\begin{equation}
    F_y = \Big\langle
    2K_y
    +2\gamma^2\sigma_{\delta}^2
    \frac{\cH{y}}{\varepsilon_y}K_z
    \Big\rangle,
\end{equation}
\begin{equation}\label{bm:Fz}
    F_z = \Big\langle2\gamma^2K_z\Big\rangle .
\end{equation}

The coefficients above include the effect of vertical dispersion through the BM formalism and provide the BM-based formulation of the same stochastic kick structure. In the simulations presented in this work, the effective coefficients are instead evaluated using Nagaitsev's formalism, as described in Sec.~\ref{AM_nag}.

\section{Convergence studies}\label{app:conv}
Choosing the appropriate number of SC kicks around the lattice of the ring and the number of macroparticles in tracking simulations is crucial to minimize computational resources without sacrificing accuracy. To address this, a convergence study was performed for two cases. These cases involved scanning different numbers of SC kicks or macroparticles in the SPS to determine the optimal parameters. The SPS lattice model, including a quadrupolar error that induces a 5--10\% beta-beating, was used for both study cases, with initial beam parameters similar to those in Sec.~\ref{sec:6}. The duration of the simulations corresponds to 100000 turns around the SPS ring.

In the first case, the number of SC kicks around the ring was varied, with the distribution initialized using 1000 macroparticles for all cases. Figure~\ref{fig:sps_conv1} presents the transverse emittances and momentum spread, normalized to their initial values, for 32, 125, 512, and 1090 SC nodes. These values correspond to approximately 0.005, 0.02, 0.075, and 0.16 SC nodes per meter around the ring, or 1.25, 5, 20, and 42 SC nodes per betatron wavelength, respectively, based on the integer part of the transverse tunes of $Q = 26$.
\begin{figure*}[!hbt]
    \centering
    \includegraphics[width=1.\textwidth]{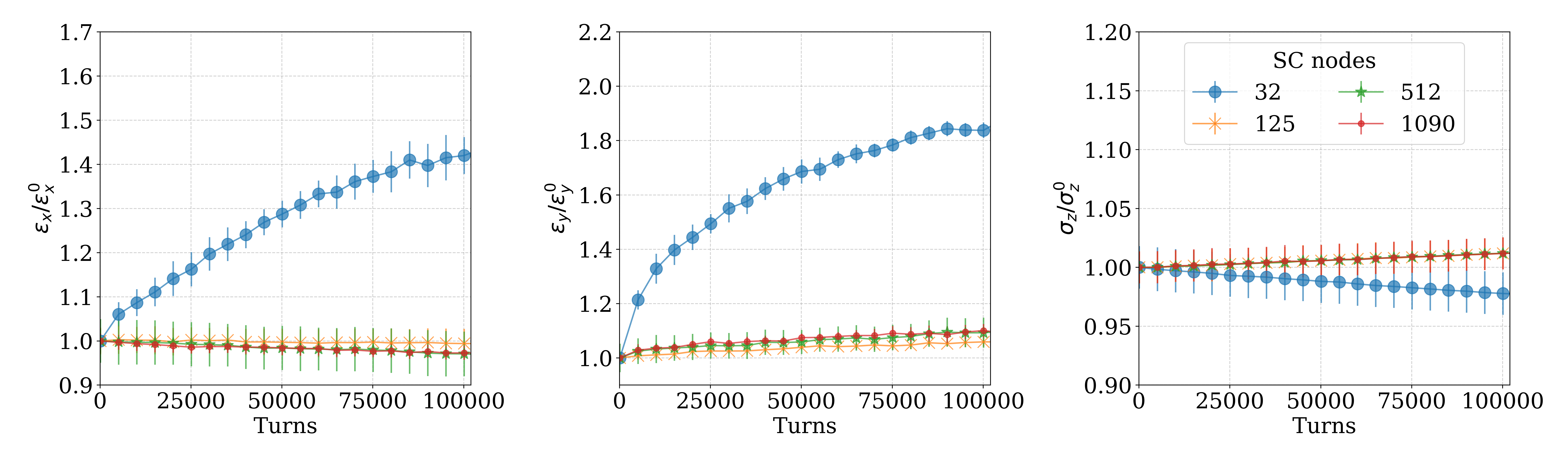}
    \caption{Evolution of the normalized horizontal emittance~(left), vertical emittance~(center) and momentum spread~(right), for 1000 macroparticles, for four different numbers of SC nodes around the SPS lattice. The error bars correspond to the standard deviation over three different simulations.}
    \label{fig:sps_conv1}
\end{figure*}

The results indicate that only the case with the smallest number of SC nodes (32) shows a significant divergence from the other three scenarios. While the case with 125 SC nodes exhibits a slight deviation from the two cases with larger numbers of SC nodes, it remains within acceptable limits. The cases with 512 and 1090 SC nodes show complete agreement with each other, and 1090 SC nodes were selected for all subsequent studies in the SPS.
\begin{figure*}[!hbt]
    \centering
    \includegraphics[width=1.\textwidth]{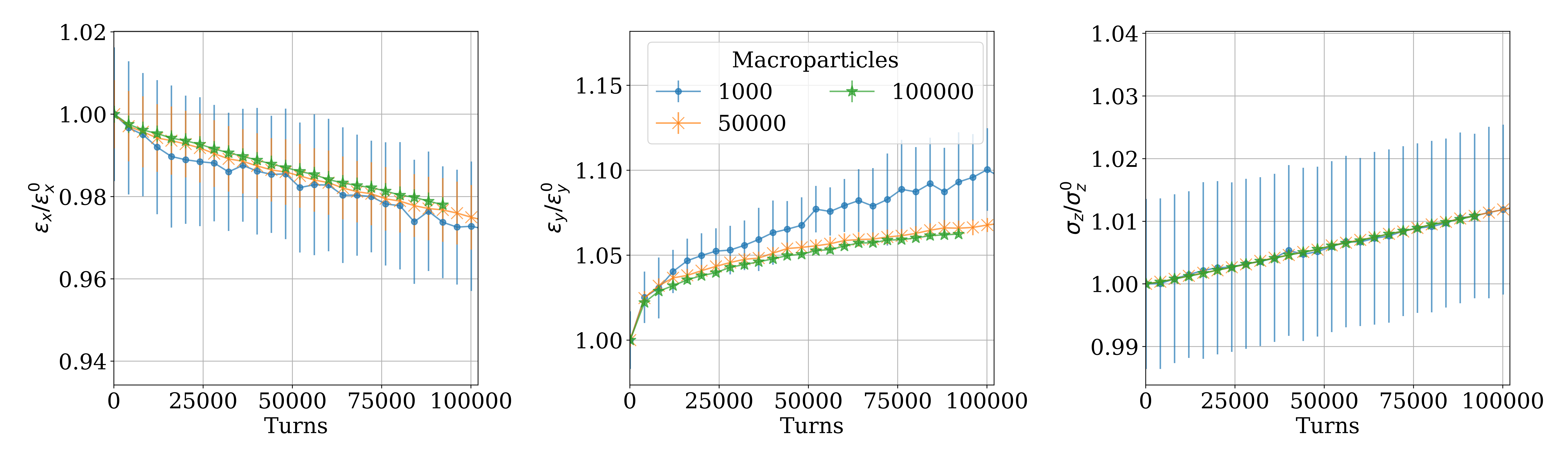}
    \caption{Evolution of the normalized horizontal emittance~(left), vertical emittance~(center) and momentum spread~(right), for 1090 SC nodes around the SPS lattice, for three different numbers of macroparticles. The error bars correspond to the standard deviation over three different simulation runs.}
    \label{fig:sps_conv2}
\end{figure*}

The required number and distribution of SC nodes depend strongly on the lattice optics and on the spatial variation of the beta and dispersion functions. In the SPS, the six-fold lattice symmetry and the large oscillations of the optical functions in the arcs make a dense sampling of SC nodes around the ring essential. Using too few SC nodes (e.g., 32) leads to under-sampling of these variations, resulting in artificial modulation of the SC tune shift and spurious emittance growth. The convergence study therefore serves to determine the minimum number of SC nodes needed to accurately reproduce the collective dynamics without introducing artificial resonances. For the SPS lattice, configurations with about 125 nodes per turn already yield satisfactory convergence, while 512 nodes provide fully stable results. The choice of 1090 equally-spaced nodes in the main simulations ensures numerical robustness under all conditions. An optimized nonuniform placement based on the local variation of the beta and dispersion functions could potentially reduce the required number of nodes; this is left for future work.

In the second study, the number of macroparticles was varied while the number of SC nodes was fixed at 1090. Figure~\ref{fig:sps_conv2} shows the transverse emittances and momentum spread, normalized to their initial values, for distributions initialized with 1000, 50000, and 100000 macroparticles. In the horizontal and longitudinal planes, all three cases agree within the statistical uncertainties, although the standard deviation over three different simulations is noticeably larger for the case of 1000 macroparticles. In the vertical plane, the two cases with larger numbers of macroparticles agree very well, while the case with 1000 macroparticles shows a 3\% difference, with error bars covering this margin. Therefore, a distribution with 1000 macroparticles is sufficient to produce accurate results while significantly reducing the simulation run time.

\section{Piwinski's invariant}\label{app:Pinv}
Following Piwinski’s original treatment of IBS~\cite{Piwinski:1974it}, one can define an approximate invariant quantity within the smooth-lattice approximation. In this framework, the optical functions are assumed to vary slowly along the ring, and derivatives of the beta functions and of the dispersion are neglected. The longitudinal degree of freedom is represented solely by the relative momentum deviation $\delta=\Delta p/p_0$.

Under these assumptions, Piwinski showed that the following quantity is approximately invariant~\cite{Martini:2016}:
\begin{equation}
\left(\frac{1}{\gamma_r^2}-\frac{1}{\gamma_\textrm{tr}^2}\right)
\langle \delta^2 \rangle
+ \frac{\varepsilon_x}{\langle \beta_x \rangle}
+ \frac{\varepsilon_y}{\langle \beta_y \rangle}
= \mathrm{const.}
\label{eq:PiwinskiEmittance}
\end{equation}
Here, $\gamma_r$ is the relativistic Lorentz factor and $\gamma_\textrm{tr}$ is the transition gamma of the ring. The transverse emittances $\varepsilon_x$ and $\varepsilon_y$ are the RMS beam emittances at a given time, while $\langle \beta_x \rangle$ and $\langle \beta_y \rangle$ denote the ring-averaged beta functions. The coefficient $(1/\gamma_r^2 - 1/\gamma_\textrm{tr}^2)$ should be interpreted as an effective global longitudinal weight arising in the smooth-lattice approximation and not as the local momentum compaction factor, particularly in a strong-focusing lattice.

Equation~\eqref{eq:PiwinskiEmittance} corresponds to the original Piwinski invariant
\begin{equation}
\left(\frac{1}{\gamma_r^2}-\frac{1}{\gamma_\textrm{tr}^2}\right)
\langle \delta^2 \rangle
+ \langle x'^2 \rangle
+ \langle y'^2 \rangle
= \mathrm{const.},
\end{equation}
rewritten in terms of RMS emittances using the smooth-lattice relations
$\langle x'^2 \rangle = \varepsilon_x / \langle \beta_x \rangle$ and
$\langle y'^2 \rangle = \varepsilon_y / \langle \beta_y \rangle$.

Although Eq.~\eqref{eq:PiwinskiEmittance} does not represent a strictly conserved quantity in a real strong-focusing lattice, it provides a useful qualitative diagnostic for understanding the redistribution of IBS-induced heating among the longitudinal and transverse degrees of freedom. In particular, the sign of the longitudinal coefficient changes across transition. Below transition ($\gamma_r < \gamma_\textrm{tr}$), all three contributions are positive and the oscillation energy remains bounded, allowing equilibrium through momentum redistribution between the three planes. Above transition ($\gamma_r > \gamma_\textrm{tr}$), the longitudinal coefficient becomes negative, indicating the absence of a bounded equilibrium in the presence of IBS alone.

To assess the applicability of this approximation, two cases from the IBS benchmarking studies of LEIR and the SPS were selected. The Piwinski invariant defined in Eq.~\eqref{eq:PiwinskiEmittance} was computed using data obtained from macroparticle simulations including the modified approximate IBS model, and from the emittance evolution predicted by the analytical IBS model of Nagaitsev~\cite{Nagaitsev}.

The first case, shown in Fig.~\ref{fig:invariant_nominal}, corresponds to the simulation presented in Fig.~\ref{fig:bench_nom}, using the nominal LEIR parameters listed in Table~\ref{tab:LEIR_params}. This is a below-transition case, where a slight reduction of the momentum spread is observed together with growth in the transverse emittances. The total Piwinski invariant obtained from macroparticle tracking simulations~(blue) and from analytical calculations~(red) shows excellent agreement. In both cases, the invariant exhibits a slow increase over time, as expected from the breakdown of the smooth-lattice approximation due to nonzero derivatives of the beta and dispersion functions in the real lattice~\cite{Piw_BM}.
\begin{figure}[!htb]
    \centering
    \includegraphics*[trim=18 0 18 15,clip,width=1.\columnwidth]{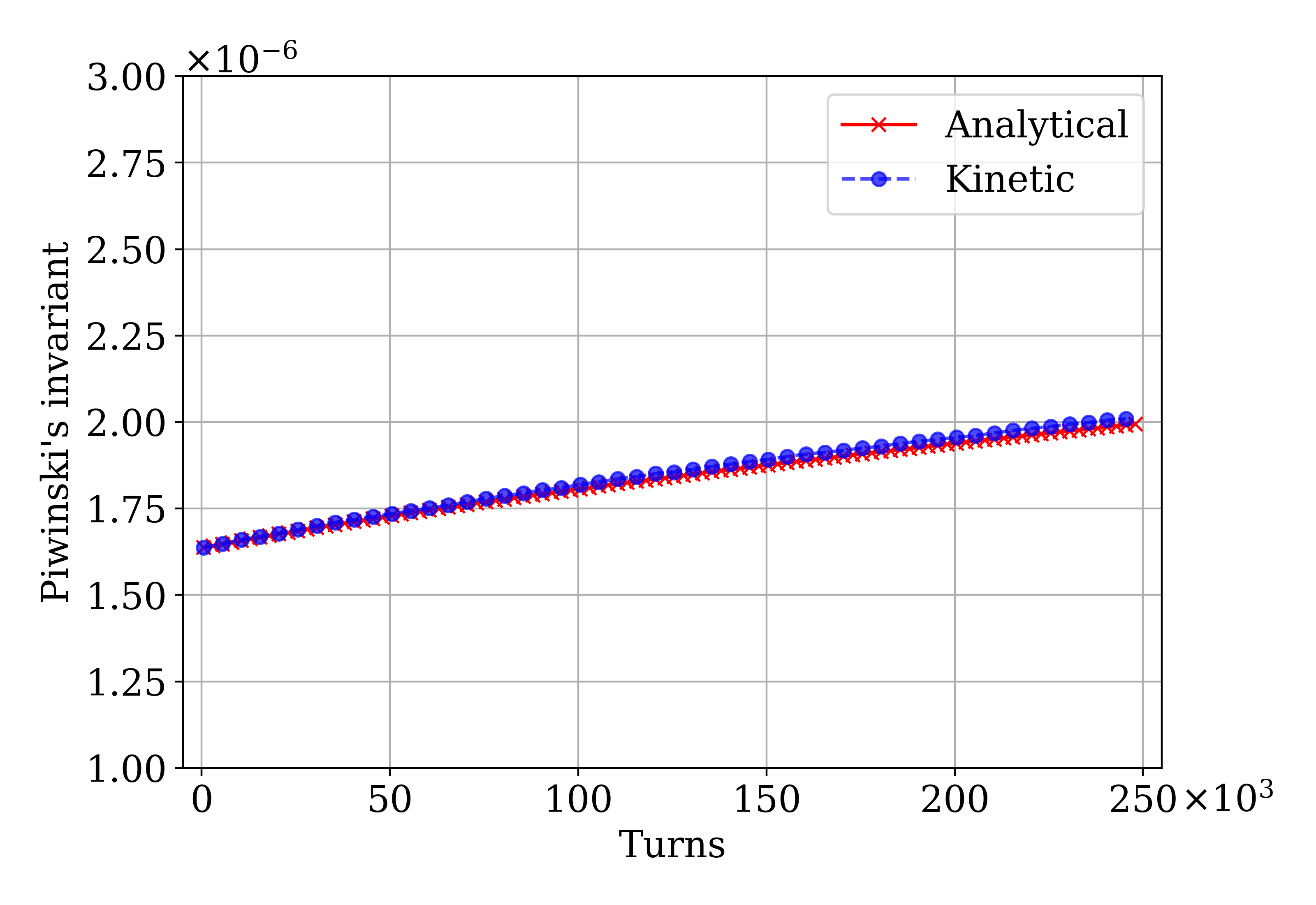}
    \caption{Piwinski's invariant, as calculated from the macroparticle tracking results using the introduced kinetic kick~(blue), and the results from analytical predictions~(red) corresponding to the case shown in Fig.~\ref{fig:bench_nom}, starting from the nominal parameters of LEIR as described in Table~\ref{tab:LEIR_params}.}
    \label{fig:invariant_nominal}
\end{figure}

The second case corresponds to the SPS simulation shown in Fig.~\ref{fig:sps_benchmarking}, where the beam is initialized with normalized transverse emittances of $\varepsilon_x^N = 1.3~\mu$m and $\varepsilon_y^N= 0.9~\mu$m, and a bunch length of $\sigma_z = 0.23$~m. Figure~\ref{fig:inv_sps1} compares the Piwinski invariant obtained from macroparticle tracking simulations~(blue) with analytical IBS predictions based on Nagaitsev’s formalism~(red), again showing good agreement. Small deviations between the two curves are observed and are attributed to the sensitivity of the invariant to statistical fluctuations in the beam second moments extracted from the macroparticle distribution; these deviations are negligible on the scale of the vertical axis. Similar to the previous case, a gradual increase of the invariant is observed, which is attributed to the nonzero derivatives of the beta and dispersion functions.
\begin{figure}[!htb]
    \centering
    \includegraphics*[trim=18 0 18 15,clip,width=1.\columnwidth]{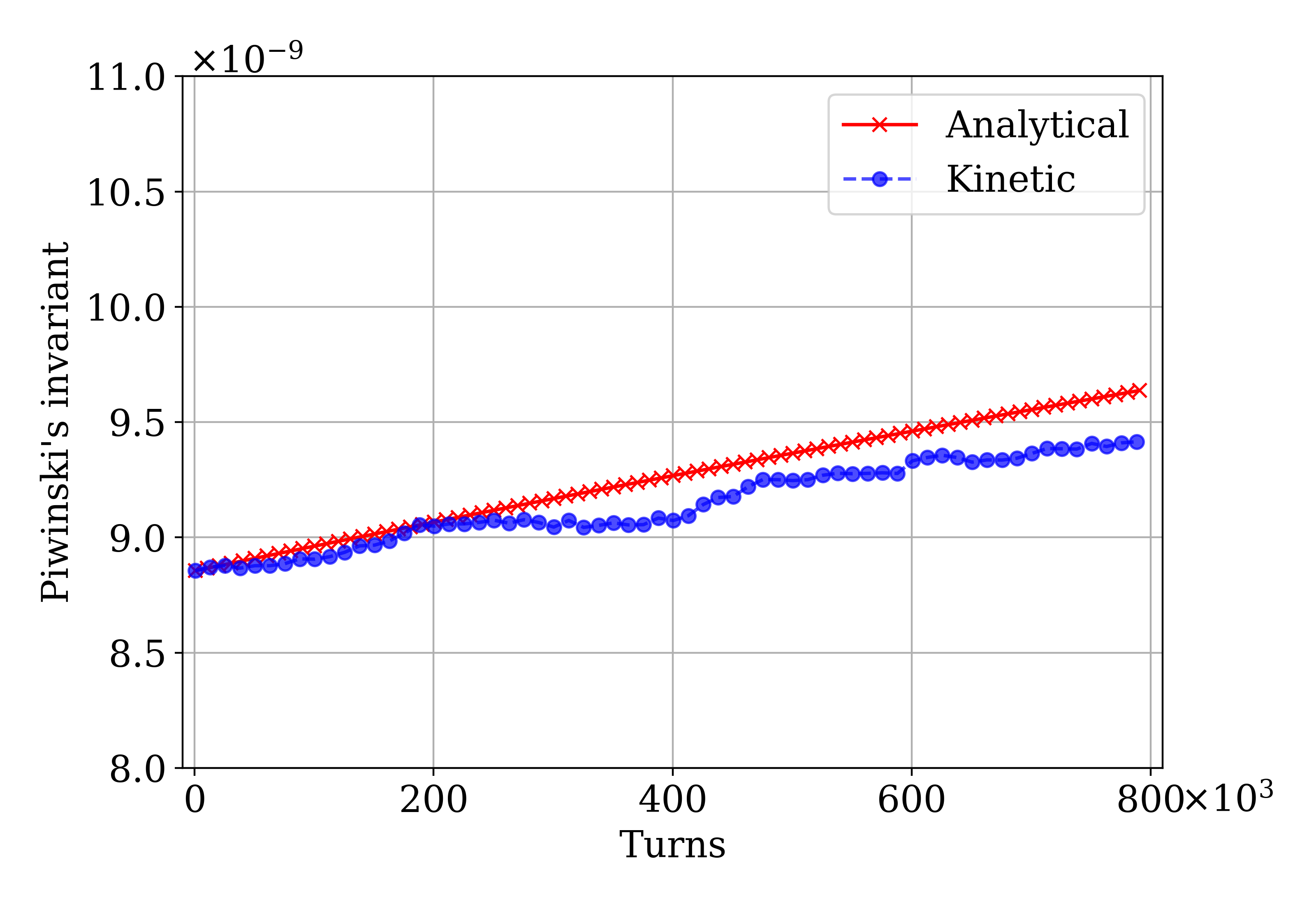}
    \caption{Piwinski's invariant, as derived from the macroparticle tracking results using the introduced kinetic kick~(blue), is compared with the analytical predictions~(red) for the case depicted in Fig.~\ref{fig:sps_benchmarking}. This case involves a Pb-ion beam initialized with normalized transverse emittances of $\eps{x}^N=1.3~\mu$m and $\eps{y}^N=0.9~\mu$m, and a bunch length of $\sigma_z=0.23$~m, in the SPS.}
    \label{fig:inv_sps1}
\end{figure}

\section{Magnetic Cycle in LEIR}\label{app:LEIRcycle}
LEIR's magnetic cycle in operation has a long injection plateau, shown in Fig.~\ref{fig:op_cycle}, as recorded by the CERN Control Center~(CCC). During this injection plateau seven beam pulses are injected from the upstream Linac3, at a kinetic energy of $4.2$~MeV/u. This is evident from the step increases in the dark blue line, which corresponds to the intensity. The seven pulses are stacked in the longitudinal phase space using an injection scheme based on combined betatron and momentum phase space painting~\cite{LHCdesignreport}.
\begin{figure}[!htb]
    \centering
    \includegraphics*[trim = 28 15 27 0, clip, width=1.\columnwidth]{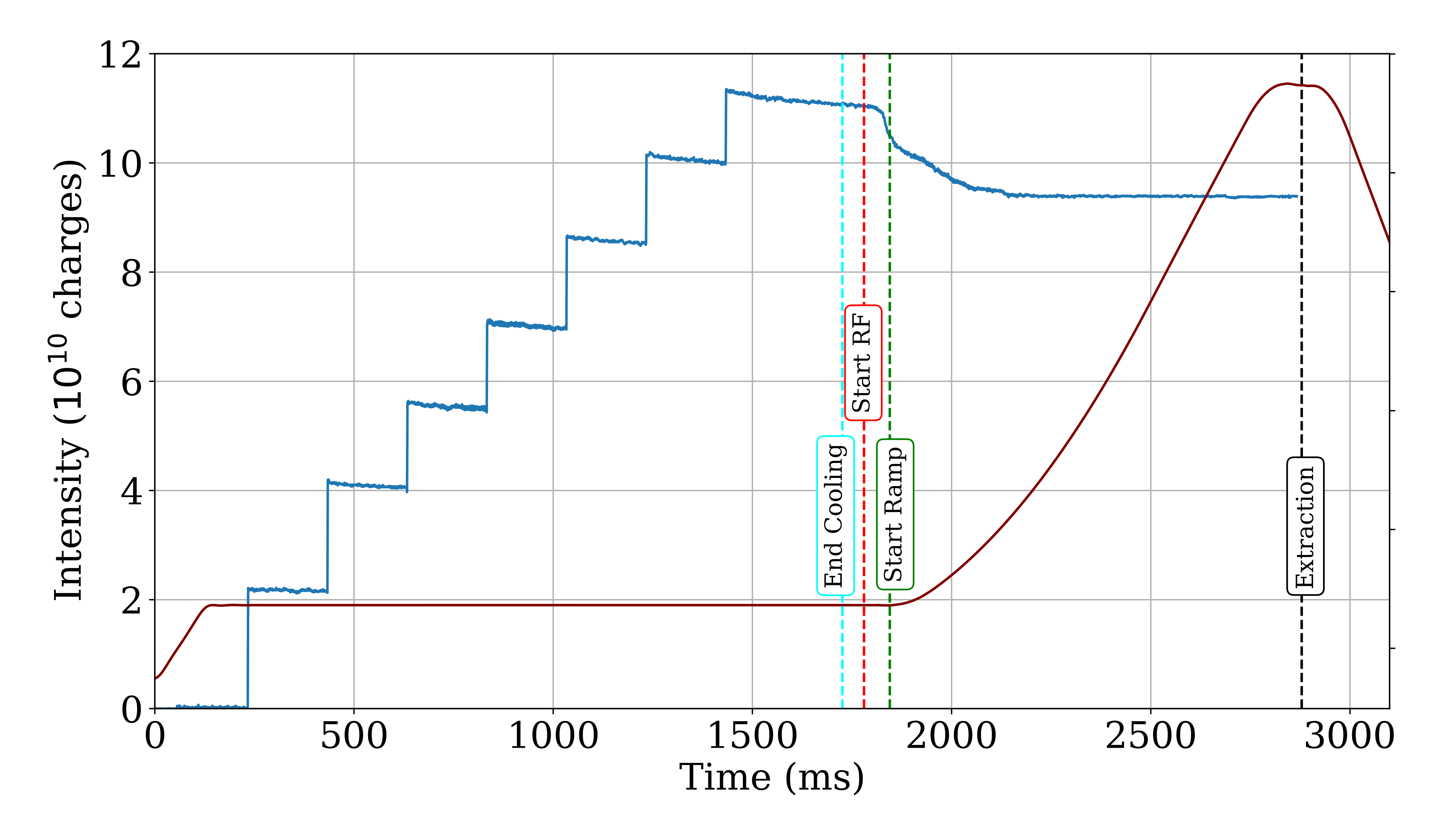}
    \caption{The operational magnetic cycle~(dark red solid line) of LEIR along with the beam intensity~(dark blue solid line) and the timings when the electron cooler is turned off~(light blue dashed line), the start of the RF-capture~(red dashed line), the start of the energy ramp~(green dashed line) and the extraction~(black dashed line).}
    \label{fig:op_cycle}
\end{figure}

Between the individual pulses, the stored coasting beam is compressed in the 6D phase space using electron cooling. By the end of the injection phase, an equilibrium is established among electron cooling, IBS-driven heating, and SC-induced emittance evolution. When the electron cooler is turned off (light blue dashed line), the RF cavity is activated~(red dashed line), capturing the coasting beam into two bunches. Consequently, the bunching increases the longitudinal line density, which enhances the SC and IBS effects, leading to particle loss. Finally, the energy ramping begins (green dashed line). Each of these steps is performed with a delay of about 60 ms between them.

The single-pulse measurements and simulations discussed in the main text were designed to isolate these effects under controlled conditions.

\section{Complementary LEIR Simulations}\label{app:LEIRsim}
Figure~\ref{fig:MD_VS_total_sim} shows the results of simulations including SC only (blue dots), IBS only (orange crosses), and the combined SC and IBS case (green stars), compared with the experimental measurements. The figure complements Fig.~\ref{fig:MD_VS_IBS_SC_notcomp} in the main text, where only the combined SC and IBS case was shown for clarity. All simulations were performed under the same conditions, including the additional lattice error introduced to reproduce the excitation of the $3Q_y=8$ resonance observed in LEIR.}
 \begin{figure}[h]
    \centering
    \includegraphics*[trim=0 25 20 20, clip, width=1.\columnwidth]{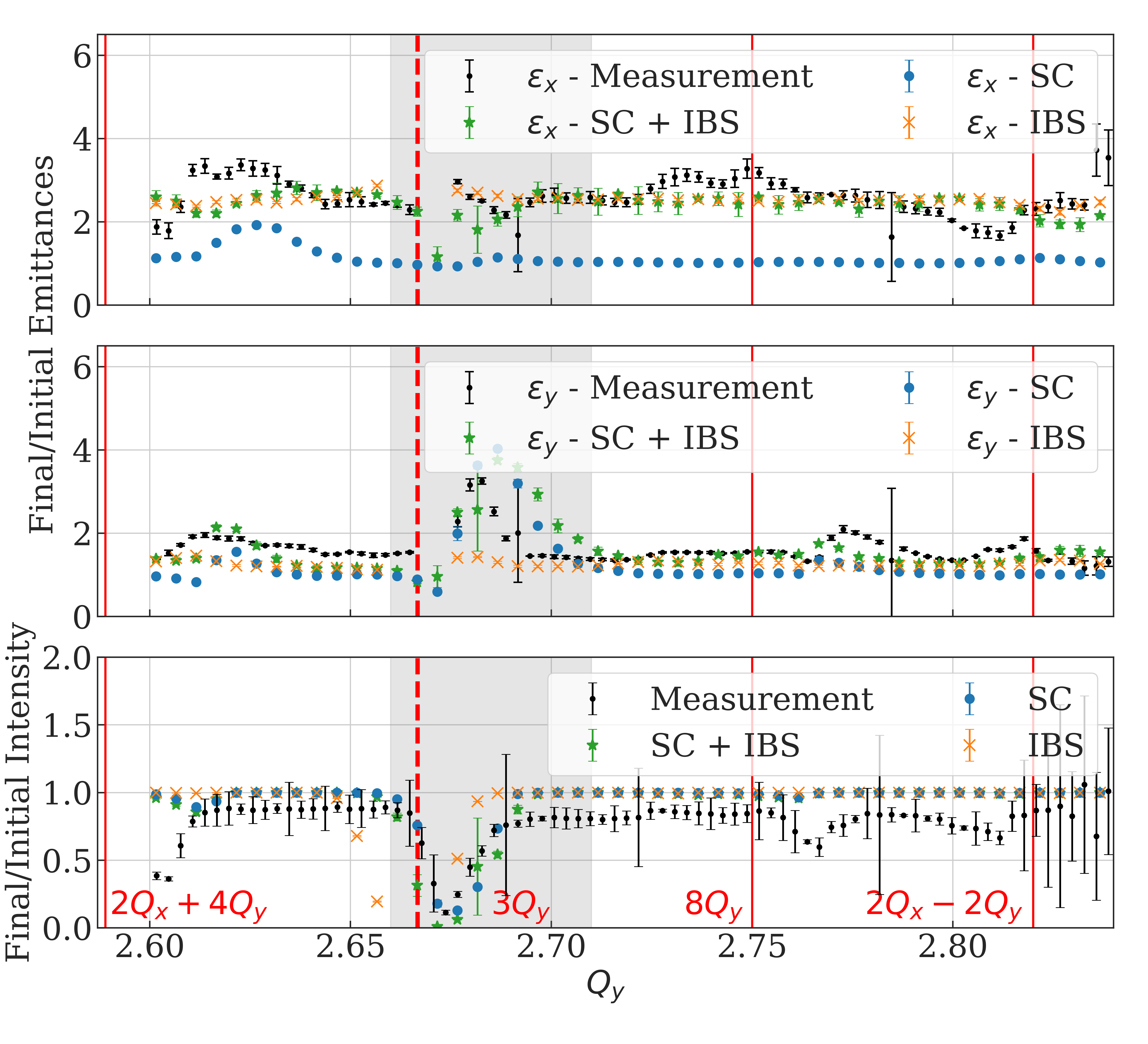}
    \caption{Measurements in LEIR~(black dots) compared with simulations including combined SC and IBS~(green stars), SC only~(blue dots), and IBS only~(orange crosses), showing the final-to-initial ratios of the horizontal~(top) and vertical~(middle) emittances, and the beam intensity~(bottom). Error bars indicate the standard deviation over three independent measurements and three independent simulation runs, respectively. This figure complements Fig.~\ref{fig:MD_VS_IBS_SC_notcomp} in the main text, where only the combined SC and IBS case was shown for clarity.}
   \label{fig:MD_VS_total_sim}
\end{figure}

As discussed in Sec.~\ref{sec:5}, the combined SC and IBS case provides the best overall agreement with the measured tune dependence of the intensity ratio and the main features of the transverse emittance behavior, whereas the individual effects alone do not reproduce the observed trends. Notably, IBS dominates the emittance growth for working points not affected by resonances.

\nocite{*}

\bibliography{references}

@PREAMBLE{
 "\providecommand{\noopsort}[1]{}" 
 # "\providecommand{\singleletter}[1]{#1}%" 
}

@article{zampetakis:clic,
    title           = "{Interplay of space charge, intrabeam scattering, and synchrotron radiation in the Compact Linear Collider damping rings}",
    author          = {Zampetakis, M. and Antoniou, F. and Asvesta, F. and Bartosik, H. and Papaphilippou, Y.},
    journal         = {Phys. Rev. Accel. Beams},
    volume          = {27},
    issue           = {6},
    pages           = {064403},
    numpages        = {20},
    year            = {2024},
    month           = {6},
    publisher       = {American Physical Society},
    doi             = {10.1103/PhysRevAccelBeams.27.064403},
    url             = {https://link.aps.org/doi/10.1103/PhysRevAccelBeams.27.064403}}

@mastersthesis{MSc_Mertens,
    author        = "Mertens, Tom",
    title         = "{Intrabeam scattering in the LHC}",
    school        = "{Porto U.}",
    year          = "2011",
    url           = "https://cds.cern.ch/record/1364596",
    note          = "Presented 17 Jun 2011"}

@article{JWei_RHIC,
    author = {Wei, J. and Fedotov, A. and Fischer, W. and Malitsky, N. and Parzen, G. and Qiang, J.},
    title = "{Intra‐beam Scattering Theory and RHIC Experiments}",
    journal = {AIP Conference Proceedings},
    volume = {773},
    number = {1},
    pages = {389-393},
    year = {2005},
    month = {06},
    issn = {0094-243X},
    doi = {10.1063/1.1949570},
    url = {https://doi.org/10.1063/1.1949570},}

@article{Fischer_RHIC,
    author      = "Fischer, W and Bai, M and Blaskiewicz, M and Brennan, J M and Cameron, P and Connolly, R and Lehrach, A and Parzen, G and Tepikian, S and Zeno, K and Van Zeijts, J",
    editor      = "Lucas, Peter W. and Webber, Sara",
    title       = "{Measurements of Intrabeam scattering growth times with gold beam below transition in RHIC}",
    reportNumber = "PAC-2001-RPAH007",
    journal     = "Conf. Proc. C",
    volume      = "0106181",
    pages       = "2857--2859",
    year        = "2001"}

@article{Bruce:ibs_kick,
    title       = "{Time evolution of the luminosity of colliding heavy-ion beams in BNL Relativistic Heavy Ion Collider and CERN Large Hadron Collider}",
    author      = {Bruce, R. and Jowett, J. M. and Blaskiewicz, M. and Fischer, W.},
    journal     = {Phys. Rev. ST Accel. Beams},
    volume      = {13},
    issue       = {9},
    pages       = {091001},
    numpages    = {16},
    year        = {2010},
    month       = {Sep},
    publisher   = {American Physical Society},
    doi         = {10.1103/PhysRevSTAB.13.091001},
    url         = {https://link.aps.org/doi/10.1103/PhysRevSTAB.13.091001}}

@article{Lebedev_1,
    author      = {Lebedev, Valeri},
    title       = "{Single and Multiple Intrabeam Scattering in Hadron Colliders}",
    journal     = {AIP Conference Proceedings},
    volume      = {773},
    number      = {1},
    pages       = {440-442},
    year        = {2005},
    month       = {06},
    issn        = {0094-243X},
    doi         = {10.1063/1.1949581},
    url         = {https://doi.org/10.1063/1.1949581},}

@article{Papadopoulou:2018uvl,
    title       = "{Impact of non-Gaussian beam profiles in the performance of hadron colliders}",
    author      = {Papadopoulou, S. and Antoniou, F. and Argyropoulos, T. and Hostettler, M. and Papaphilippou, Y. and Trad, G.},
    journal     = {Phys. Rev. Accel. Beams},
    volume      = {23},
    issue       = {10},
    pages       = {101004},
    numpages    = {21},
    year        = {2020},
    month       = {Oct},
    publisher   = {American Physical Society},
    doi         = {10.1103/PhysRevAccelBeams.23.101004},
    url         = {https://link.aps.org/doi/10.1103/PhysRevAccelBeams.23.101004}}

@phdthesis{PhD_Papadopoulou,
    author      = "Papadopoulou, Parthena Stefania",
    title       = "{Bunch characteristics evolution for lepton and hadron rings under the influence of the Intra-beam scattering effect}",
    reportNumber= "CERN-THESIS-2019-405",
    school      = "University of Crete",
    year        = "2019"}

@techreport{Lebedev_2,
    author      = "Lebedev, V. A. and Lobach, I. and Romanov, A. and Valishev, A.",
    title       = "{Report on Single and Multiple Intrabeam Scattering Measurements in IOTA Ring in Fermilab}",
    institution = "Fermi National Accelerator Lab., Batavia, IL (United States)",
    reportNumber= "FERMILAB-TM-2750-AD, Beams-doc-8837",
    doi         = "10.2172/1764146",
    year        = "2020"}

@phdthesis{PhD_Antoniou,
    author      = "Antoniou, Fanouria",
    title       = "{Optics design of Intrabeam Scattering dominated damping rings}",
    school      = "{National Technical University of Athens}",
    year        = "2012",
    url         = "http://cds.cern.ch/record/1666863"}

@inproceedings{Antoniou:SLS,
    author      = "Antoniou, F and Papaphilippou, Y and Aiba, M and Boege, M and Milas, N and Streun, A and Demma, T",
    title       = "{Intrabeam scattering studies at the Swiss light source}",
    booktitle   = {Proc. 2nd International Particle Accelerator Conference (IPAC'12)},
    pages       = {1951--1953},
    paper       = {TUPPR057},
    month       = "May",
    year        = "2012",
    address     = {New Orleans, Louisiana, USA},
    url         = "https://cds.cern.ch/record/1464110"}

@techreport{Bane_2,
    title       = "{Intra-Beam Scattering, Impedance, and Instabilities in Ultimate Storage Rings}",
    author      = {Karl L. F. Bane},
    abstractNote = {We have investigated collective effects in an ultimate storage ring, i.e. one with diffraction limited emittances in both planes, using PEP-X as an example. In an ultimate ring intra-beam scattering (IBS) sets the limit of current that can be stored. In PEP-X, a 4.5 GeV ring running round beams at 200 mA in 3300 bunches, IBS doubles the emittances to 11.5 pm at the design current. The Touschek lifetime is 11 hours. Impedance driven collective effects tend not to be important since the beam current is relatively low. We have investigated collective effects in PEP-X, an ultimate storage ring, i.e. one with diffraction limited emittances (at one angstrom wavelength) in both planes. In an ultimate ring intra-beam scattering (IBS) sets the limit of current that can be stored. In PEP-X, IBS doubles the emittances to 11.5 pm at the design current of 200 mA, assuming round beams. The Touschek lifetime is quite large in PEP-X, 11.6 hours, and - near the operating point - increases with decreasing emittance. It is, however, a very sensitive function of momentum acceptance. In an ultimate ring like PEP-X impedance driven collective effects tend not to be important since the beam current is relatively low. Before ultimate PEP-X can be realized, the question of how to run a machine with round beams needs serious study. For example, in this report we assumed that the vertical emittance is coupling dominated. It may turn out that using vertical dispersion is a preferable way to generate round beams. The choice will affect IBS and the Touschek effect.},
    reportNumber = "SLAC-PUB-14892",
    url         = {https://www.osti.gov/biblio/1037597}, 
    institution = "SLAC National Accelerator Lab., Menlo Park, CA (United States)",
    place       = {United States},
    year        = {2012},
    month       = {3}}

@article{Bane_KEK,
    author 		  = {K. L. F. Bane and H. Hayano and K. Kubo and T. Naito and T. Okugi and J. Urakawa},
    title 		  = "{Intrabeam scattering analysis of measurements at {KEK}'s Accelerator Test Facility damping ring}",
    journal 	  = {Physical Review Special Topics - Accelerators and Beams},
    year 		  = "2002",
    month 		  = {Aug},
    publisher 	  = {American Physical Society ({APS})},
    volume 		  = {5},
    number 		  = {8},
    url 		  = {https://doi.org/10.1103\%2Fphysrevstab.5.084403},
    doi 		  = {10.1103/physrevstab.5.084403}}

@article{Kubo_KEK,
    title       = "{Intrabeam scattering formulas for high energy beams}",
    author      = {Kubo, Kiyoshi and Mtingwa, Sekazi K. and Wolski, Andrzej},
    journal     = {Phys. Rev. ST Accel. Beams},
    volume      = {8},
    issue       = {8},
    pages       = {081001},
    numpages    = {8},
    year        = {2005},
    month       = {Aug},
    publisher   = {American Physical Society},
    doi         = {10.1103/PhysRevSTAB.8.081001},
    url         = {https://link.aps.org/doi/10.1103/PhysRevSTAB.8.081001}}

@techreport{Huang,
    author      = "Huang, Z.",
    title       = "{Intrabeam scattering in an X-ray FEL driver}",
    reportNumber = "SLAC-TN-05-026, LCLS-TN-02-8",
    month       = "11",
    institution = "SLAC National Accelerator Lab., Menlo Park, CA (United States)",
    year        = "2002"}

@inproceedings{Xiao_2010,
    author      = "Xiao, Aimin and Borland, Michael",
    title       = "{Intrabeam Scattering Effect Calculated for a Non-Gaussian Distributed Linac Beam}",
    booktitle   = "{Proc. PAC'09}",
    eid         = {TH5PFP038},
    pages       = "3281--3283",
    venue       = {Vancouver, BC, Canada},
    year        = "2010"}

@article{DiMitri_2020,
    author      = {S Di Mitri and G Perosa and A Brynes and I Setija and S Spampinati and P H Williams and A Wolski and E Allaria and S Brussaard and L Giannessi and G Penco and P R Rebernik and M Trovò},
    title       = "{Experimental evidence of intrabeam scattering in a free-electron laser driver}",
    doi         = {10.1088/1367-2630/aba572},
    url         = {https://dx.doi.org/10.1088/1367-2630/aba572},
    year        = {2020},
    month       = {aug},
    publisher   = {IOP Publishing},
    volume      = {22},
    number      = {8},
    pages       = {083053},
    journal     = {New Journal of Physics}}

@article{Hunt_Carli,
    title       = {Emittance measurements in low energy ion storage rings},
    journal     = {Nuclear Instruments and Methods in Physics Research Section A: Accelerators, Spectrometers, Detectors and Associated Equipment},
    volume      = {896},
    pages       = {139-151},
    year        = {2018},
    issn        = {0168-9002},
    doi         = {https://doi.org/10.1016/j.nima.2018.04.018},
    url         = {https://www.sciencedirect.com/science/article/pii/S0168900218305035},
    author      = {J.R. Hunt and C. Carli and J. Resta-López and C.P. Welsch}}

@inproceedings{RHIC,
    title       = "{Beam lifetime and limitations during low-energy RHIC operation}",
    author      = {Alexei Fedotov and Mei Bai and Michael M. Blaskiewicz and W. Fischer and Dmitry Kayran and Christoph Montag and Todd Satogata and Steven Tepikian and Gang Wang},
    booktitle   = {Proc. PAC'11},
    pages       = {2285--2287},
    eid         = {THP081},
    venue       = {New York, NY, United States},
    series      = {Particle Accelerator Conference},
    year        = {2011}}

@article{Siggel,
    author      = "Siggel-King, M. R. F. and Welsch, C. P. and Papash, A. I. and Smirnov, A. V.",
    editor      = "Petit-Jean-Genaz, Christine",
    title       = "{Long Term Beam Dynamics in Ultra-Low Energy Storage Rings}",
    reportNumber = "IPAC-2011-WEPC064",
    journal     = "Conf. Proc. C",
    volume      = "110904",
    pages       = "2166--2168",
    year        = "2011"}

@article{Franchetti:2003aa,
    author      = "Franchetti, G. and Hofmann, I. and Giovannozzi, M. and Martini, M. and Metral, E.",
    title       = "{Space charge and octupole driven resonance trapping observed at the CERN Proton Synchrotron}",
    doi         = "10.1103/PhysRevSTAB.6.124201",
    journal     = "Phys. Rev. ST Accel. Beams",
    volume      = "6",
    pages       = "124201",
    year        = "2003"}

@article{Metral:2006qw,
    title       = "{Observation of octupole driven resonance phenomena with space charge at the CERN Proton Synchrotron}",
    journal     = {Nuclear Instruments and Methods in Physics Research Section A: Accelerators, Spectrometers, Detectors and Associated Equipment},
    volume      = {561},
    number      = {2},
    pages       = {257-265},
    year        = {2006},
    note        = {Proceedings of the Workshop on High Intensity Beam Dynamics},
    issn        = {0168-9002},
    doi         = {https://doi.org/10.1016/j.nima.2006.01.029},
    url         = {https://www.sciencedirect.com/science/article/pii/S0168900206000428},
    author      = {E. M\'etral and G. Franchetti and M. Giovannozzi and I. Hofmann and M. Martini and R. Steerenberg}}

@article{Franchetti:2010zz,
    title       = {Experiment on space charge driven nonlinear resonance crossing in an ion synchrotron},
    author      = {Franchetti, G. and Chorniy, O. and Hofmann, I. and Bayer, W. and Becker, F. and Forck, P. and Giacomini, T. and Kirk, M. and Mohite, T. and Omet, C. and Parfenova, A. and Sch\"utt, P.},
    journal     = {Phys. Rev. ST Accel. Beams},
    volume      = {13},
    issue       = {11},
    pages       = {114203},
    numpages    = {22},
    year        = {2010},
    month       = {Nov},
    publisher   = {American Physical Society},
    doi         = {10.1103/PhysRevSTAB.13.114203},
    url         = {https://link.aps.org/doi/10.1103/PhysRevSTAB.13.114203}}

@article{Franchetti:2017aa,
    title       = {Space charge effects on the third order coupled resonance},
    author      = {Franchetti, Giuliano and Gilardoni, Simone and Huschauer, Alexander and Schmidt, Frank and Wasef, Raymond},
    journal     = {Phys. Rev. Accel. Beams},
    volume      = {20},
    issue       = {8},
    pages       = {081006},
    numpages    = {16},
    year        = {2017},
    month       = {Aug},
    publisher   = {American Physical Society},
    doi         = {10.1103/PhysRevAccelBeams.20.081006},
    url         = {https://link.aps.org/doi/10.1103/PhysRevAccelBeams.20.081006}}

@article{Asvesta:ps,
    title       = "{Identification and characterization of high order incoherent space charge driven structure resonances in the CERN Proton Synchrotron}",
    author      = {Asvesta, F. and Bartosik, H. and Gilardoni, S. and Huschauer, A. and Machida, S. and Papaphilippou, Y. and Wasef, R.},
    journal     = {Phys. Rev. Accel. Beams},
    volume      = {23},
    issue       = {9},
    pages       = {091001},
    numpages    = {20},
    year        = {2020},
    month       = {Sep},
    publisher   = {American Physical Society},
    doi         = {10.1103/PhysRevAccelBeams.23.091001},
    url         = {https://link.aps.org/doi/10.1103/PhysRevAccelBeams.23.091001}}

@article{SaaHernandez:2018zqv,
    author      = "Sa\'a Hern\'andez, Angela and Bartosik, Hannes and Biancacci, Nicolo and Hirlaender, Simon and Huschauer, Alexander and Moreno Garcia, Daniel",
    title       = "{Space Charge Studies on LEIR}",
    journal     = "J. Phys. Conf. Ser.",
    volume      = "1067",
    number      = "6",
    pages       = "062020",
    year        = "2018",
    doi         = "10.18429/JACoW-IPAC2018-THPAF055"}

@article{Fermi_res,
    title       = "{High intensity space charge effects on slip stacked beam in the Fermilab Recycler}",
    author      = {Ainsworth, R. and Adamson, P. and Amundson, J. and Kourbanis, I. and Lu, Q. and Stern, E.},
    journal     = {Phys. Rev. Accel. Beams},
    volume      = {22},
    issue       = {2},
    pages       = {020404},
    numpages    = {9},
    year        = {2019},
    month       = {Feb},
    publisher   = {American Physical Society},
    doi         = {10.1103/PhysRevAccelBeams.22.020404},
    url         = {https://link.aps.org/doi/10.1103/PhysRevAccelBeams.22.020404}}

@inproceedings{Fermi_IOTA,
    author      = {V.D. Shiltsev},
    title       = {{Space-Charge and Other Effects in Fermilab Booster and IOTA Rings’ Ionization Profile Monitors}},
    booktitle   = {Proc. IBIC'21},
    pages       = {193--197},
    eid         = {TUPP05},
    venue       = {Pohang, Rep. of Korea},
    series      = {International Beam Instrumentation Conference},
    number      = {10},
    publisher   = {JACoW Publishing, Geneva, Switzerland},
    month       = {10},
    year        = {2021},
    issn        = {2673-5350},
    isbn        = {978-3-95450-230-1},
    doi         = {10.18429/JACoW-IBIC2021-TUPP05},
    url         = {https://jacow.org/ibic2021/papers/tupp05.pdf},
    note        = {https://doi.org/10.18429/JACoW-IBIC2021-TUPP05}}

@article{Fermi_boost,
    author      = {Yu. Alexahin and V. Kapin},
    title       = "{On possibility of space-charge compensation in the Fermilab Booster with multiple electron columns}",
    doi         = {10.1088/1748-0221/16/03/P03049},
    url         = {https://dx.doi.org/10.1088/1748-0221/16/03/P03049},
    year        = {2021},
    month       = {mar},
    publisher   = {IOP Publishing},
    volume      = {16},
    number      = {03},
    pages       = {P03049},
    journal     = {Journal of Instrumentation}}

@inproceedings{Fermi_highI,
    author      = "Chung, M. and Shiltsev, V. and Prost, L.",
    title       = "{Space-Charge Compensation for High-Intensity Linear and Circular Accelerators at Fermilab}",
    booktitle   = "{1st North American Particle Accelerator Conference}",
    reportNumber = "FERMILAB-CONF-13-412-APC",
    address     = {Pasadena, CA, USA},
    month       = "9",
    year        = "2013"}

@inproceedings{SC_RHIC,
    author      = {Dell, G.F. and Peggs, S.},
    title       = "{Simulation of the space charge effect in RHIC}", 
    booktitle   = {Proceedings Particle Accelerator Conference}, 
    address     = {Dallas, TX, USA},
    year        = {1995},
    volume      = {4},
    pages       = {2327-2329},
    doi         = {10.1109/PAC.1995.505541}}

@article{SC_eRHIC,
    title       = {Compensating tune spread induced by space charge in bunched beams},
    author      = {Litvinenko, Vladimir N. and Wang, Gang},
    journal     = {Phys. Rev. ST Accel. Beams},
    volume      = {17},
    issue       = {11},
    pages       = {114401},
    numpages    = {24},
    year        = {2014},
    month       = {Nov},
    publisher   = {American Physical Society},
    doi         = {10.1103/PhysRevSTAB.17.114401},
    url         = {https://link.aps.org/doi/10.1103/PhysRevSTAB.17.114401}}

@article{Venturini:2006rp,
    author      = "Venturini, M. and Oide, K. and Wolski, A.",
    editor      = "Prior, Christopher",
    title       = "{Space charge and equilibrium emittances in damping rings}",
    reportNumber= "LBNL-59509",
    journal     = "Conf. Proc. C",
    volume      = "060626",
    pages       = "882--884",
    year        = "2006"}

@techreport{Venturini:2006ilc,
    author      = "Venturini, M. and Oide, K.",
    title       = "{Direct space charge effects on the ILC damping rings: Task force report}",
    institution = "Lawrence Berkeley National Laboratory (LBNL)",
    address     = "Berkeley, CA (United States)",
    month       = "2",
    year        = "2006",
    reportNumber= "LBNL-59511",
    url         = "https://cds.cern.ch/record/2749453",
    doi         = "10.2172/889308"}

@techreport{Venturini:2007ler,
    title       = {Space-Charge Effects in the Super B-Factory LER},
    author      = {Venturini, Marco},
    institution = "Lawrence Berkeley National Laboratory (LBNL)",
    address     = "Berkeley, CA (United States)",
    abstractNote = {Space-charge effects in the low-energy ring of the proposedSuper-B Factory are studied using a weak-strong model of dynamics asimplemented in the code Marylie/Impact (MLI). The impact of space chargeappears noticeable but our results suggest the existence of workableregions of the tune space where the design emittance is minimallyaffected. However, additional studies are recommended to fullysubstantiate this conclusion.},
    doi         = {10.2172/923014},
    url         = {https://www.osti.gov/biblio/923014},
    year        = {2007},
    month       = {1}}

@article{Xiao:2007ilc,
    author      = "Xiao, A. and Borland, Michael and Emery, L. and Wang, Y. and Ng, K. Y.",
    editor      = "Petit-Jean-Genaz, C.",
    title       = "{Direct Space Charge Calculation in Elegant and Its Application to the ILC Damping Ring}",
    reportNumber = "FERMILAB-CONF-07-702-AD, PAC07-THPAN099",
    doi         = "10.2172/921985",
    journal     = "Conf. Proc. C",
    volume      = "070625",
    pages       = "3456",
    year        = "2007"}

@article{mocac,
    author      = {Alekseev, N. and Bolshakov, A. and Mustafin, E. and Zenkevich, P.},
    title       = "{Numerical code for Monte-Carlo simulation of ion storage}",
    journal     = {AIP Conference Proceedings},
    volume      = {480},
    number      = {1},
    pages       = {31-41},
    year        = {1999},
    month       = {06},
    issn        = {0094-243X},
    doi         = {10.1063/1.59502},
    url         = {https://doi.org/10.1063/1.59502},
    eprint      = {https://pubs.aip.org/aip/acp/article-pdf/480/1/31/11636089/31\_1\_online.pdf}}

@inproceedings{Vivoli:1,
    author       = "Vivoli, Alessandro and Martini, Michel",
    title        = "{Intra-Beam Scattering in the CLIC Damping Rings}",
    reportNumber = "IPAC-2010-WEPE090",
    month        = {05},
    year         = {2010},
    booktitle    = {Proc. IPAC'10},
    pages        = {3557--3559},
    eid          = {WEPE090},
    venue        = {Kyoto, Japan},
    series       = {International Particle Accelerator Conference},
    number       = {1}}

@inproceedings{osti_1,
    author 		= {Biagini, M. and Boscolo, Manuela and Demma, T. and Chao, A. and Bane, Karl and Pivi, M.},
    title 		= {Multiparticle Simulation of Intrabeam Scattering for SuperB},
    reportNumber= {IPAC-2011-WEPC105},
    month       = {09},
    year        = {2011},
    booktitle   = {Proc. IPAC'11},
    pages       = {2259--2261},
    eid         = {WEPC105},
    venue       = {San Sebastian, Spain},
    series      = {International Particle Accelerator Conference},
    number      = {2}}

@article{Pivi:1,
    author 		  = "Pivi, M. T. F.",
    editor 		  = "Petit-Jean-Genaz, C.",
    title 		  = "{CMAD: A new self-consistent parallel code to simulate the electron cloud build-up and instabilities}",
    reportNumber  = "PAC07-THPAS066, SLAC-PUB-12970",
    doi 		  = "10.1109/PAC.2007.4440517",
    journal 	  = "Conf. Proc. C",
    volume 		  = "070625",
    pages 		  = "3636",
    year 		  = "2007"}

@article{Sonnad:1,
    author 		  = "Sonnad, K. G. and Antoniou, F. and Papaphilippou, Y. and Li, K. S. B. and Boscolo, M. and Demma, T. and Chao, A. and Rivetta, C. H. and Pivi, M. T. F.",
    editor 		  = "Suller, Vic",
    title 		  = "{Multi-Particle Simulation Codes Implementation to Include Models of a Novel Single-bunch Feedback System and Intra-beam Scattering}",
    reportNumber  = "IPAC-2012-WEPPR091, SLAC-PUB-16030",
    journal 	  = "Conf. Proc. C",
    volume 	 	  = "1205201",
    pages 		  = "3147--3149",
    year 		  = "2012"}

@article{Zenkevich:0,
    author      = {P. Zenkevich and O. Boine-Frankenheim and A. Bolshakov},
    title       = {A new algorithm for the kinetic analysis of intra-beam scattering in storage rings},
    journal     = {Nuclear Instruments and Methods in Physics Research Section A: Accelerators, Spectrometers, Detectors and Associated Equipment},
    volume      = {561},
    number 		= {2},
    pages 		= {284-288},
    year  	    = {2006},
    issn 		= {0168-9002},
    doi 		= {https://doi.org/10.1016/j.nima.2006.01.013},
    url	 		= {https://www.sciencedirect.com/science/article/pii/S0168900206000465}}

@article{Zenkevich:1,
    author 		= "Zenkevich, P. and Bolshakov, A. and Boine-Frankenheim, O.",
    title 		= "{Kinetic effects in multiple intra-beam scattering}",
    journal 	= "AIP Conf. Proc.",
    volume 		= "773",
    number 		= "1",
    pages 		= "425--429",
    year 		= "2005",
    doi 		= "10.1063/1.1949577"}

@article{Zenkevich:2,
    author      = {Bolshakov, A. and Zenkevich, P.},
    year        = {2012},
    month       = {07},
    pages       = {170--177},
    title       = {Numerical modeling of intra-beam scattering of particles in ring accelerators and storage rings by approximate solution of the fokker–planck equation in the space of invariants},
    volume      = {112},
    journal     = {Atomic Energy},
    doi         = {10.1007/s10512-012-9539-0}}

@Inbook{Risken1996,
    author      = "Risken, Hannes",
    title       = "Fokker-Planck Equation",
    bookTitle   = "The Fokker-Planck Equation: Methods of Solution and Applications",
    year        = "1996",
    publisher   = "Springer Berlin Heidelberg",
    address     = "Berlin, Heidelberg",
    pages       = "63--95",
    isbn        = "978-3-642-61544-3",
    doi         = "10.1007/978-3-642-61544-3_4",
    url         = "https://doi.org/10.1007/978-3-642-61544-3_4"}

@article{Nagaitsev,
    title       = {Intrabeam scattering formulas for fast numerical evaluation},
    author      = {Nagaitsev, Sergei},
    journal     = {Phys. Rev. ST Accel. Beams},
    volume      = {8},
    issue       = {6},
    pages       = {064403},
    numpages    = {4},
    year        = {2005},
    month       = {Jun},
    publisher   = {American Physical Society},
    doi         = {10.1103/PhysRevSTAB.8.064403},
    url         = {https://link.aps.org/doi/10.1103/PhysRevSTAB.8.064403}}

@article{Bjorken:0,
    author      = "Bjorken, James D and Mtingwa, S K",
    title       = "{Intrabeam scattering}",
    journal     = "Part. Accel.",
    reportNumber= "FERMILAB-PUB-82-47-THY",
    volume      = "13",
    pages       = "115-143. 63 p",
    month       = "Jul",
    year        = "1982",
    url         = "https://cds.cern.ch/record/140304"}

@article{Conte:1985,
    author      = "Conte, M. and Martini, M.",
    title       = "{Intrabeam Scattering in the CERN Anti-Proton Accumulator}",
    journal     = "Part. Accel.",
    volume      = "17",
    pages       = "1--10",
    year        = "1985"}

@techreport{Antoniou:mad,
    author        = "Antoniou, F and Zimmermann, F",
    title         = "{Revision of Intrabeam Scattering with Non-Ultrarelativistic Corrections and Vertical Dispersion for MAD-X}",
    institution   = "CERN",
    reportNumber  = "CERN-ATS-2012-066",
    address       = "Geneva, Switzerland",
    year          = "2012",
    url           = "https://cds.cern.ch/record/1445924"}

@article{Carlson,
    author		  = "{Carlson, B. C.}",
    title 		  = "{Three Improvements in Reduction and Computation of Elliptic Integrals}",
    journal		  = "Journal of research of the National Institute of Standards and Technology",
    year		  = "2002",
    volume		  = {105(5)},
    pages		  = {413-418},
    url 		  = "https://doi.org/10.6028/jres.107.034"}

@inproceedings{Piwinski:1974it,
    author      = "Piwinski, A.",
    title       = "{Intra-beam-Scattering}",
    booktitle   = "{9th International Conference on High-Energy Accelerators}",
    reportNumber = "DESY-H1-73-3",
    pages       = "405--409",
    year        = "1974",
    venue      = {SLAC, Stanford, CA, USA}}

@techreport{PTC,
    author      = "Schmidt, F and Forest, E and McIntosh, E",
    title       = "{Introduction to the polymorphic tracking code: Fibre bundles, polymorphic Taylor types and ''Exact tracking''}",
    institution = "CERN",
    address     = "Geneva, Switzerland",
    reportNumber= "CERN-SL-2002-044-AP, KEK-REPORT-2002-3",
    month       = "Jul",
    year        = "2002",
    url         = "http://cds.cern.ch/record/573082",}

@article{pyorbit,
    author      = {Andrei Shishlo and Sarah Cousineau and Jeffrey Holmes and Timofey Gorlov},
    title       = "{The Particle Accelerator Simulation Code PyORBIT}",
    journal     = {Procedia Computer Science},
    volume      = {51},
    pages       = {1272-1281},
    year        = {2015},
    note        = {International Conference On Computational Science, ICCS 2015},
    issn        = {1877-0509},
    doi         = {https://doi.org/10.1016/j.procs.2015.05.312},
    url         = {https://www.sciencedirect.com/science/article/pii/S1877050915011205}}

@inproceedings{shishlo:pyorbit,
    author      = {Shishlo, A and Cousineau, Sarah and Danilov, V and Galambos, J and Henderson, Skye and Holmes, J.A.},
    title       = "{The ORBIT simulation code: benchmarking and applications}",
    booktitle   = {Proc. ICAP 2006},
    year        = {2006},
    eid         = {MOA2IS01},
    pages       = {53--58},
    address     = {Chamonix, France},
    month       = {10}}

@phdthesis{PhD_Asvesta,
    author      = "Asvesta, Foteini",
    title       = "{Space charge and lattice driven  resonances at the CERN injectors.}",
    school      = "{National Technical University of Athens}",
    year        = "2020",
    url         = "https://cds.cern.ch/record/2771289"}

@inproceedings{Schmidt:2016izr,
    author      = "Schmidt, Frank and Alexahin, Yuri and Amundson, James and Bartosik, Hannes and Franchetti, Giuliano and Holmes, Jeffrey and Huschauer, Alexander and Kapin, Valery and Oeftiger, Adrian and Stern, Eric and Titze, Malte",
    title       = "{Code bench-marking for long-term tracking and adaptive algorithms}",
    booktitle   = {Proc. of ICFA Advanced Beam Dynamics Workshop on High-Intensity and High-Brightness Hadron Beams (HB'16)},
    pages       = {357--361},
    address     = {Malmö, Sweden},
    year        = {2016},
    month       = {7},
    doi         = {doi:10.18429/JACoW-HB2016-WEAM1X01},
    url         = {http://jacow.org/hb2016/papers/weam1x01.pdf}}

@misc{Li:pyorbit,
    author      = {Li, Runze and Sen, Tanaji and Ostiguy, Jean-Francois},
    title       = "{Modeling Transverse Space Charge effects in IOTA with pyORBIT}",
    year        = {2021},
    eprint      = {2106.03327},
    archivePrefix = {arXiv},
    primaryClass  = {physics.acc-ph},
    reportNumber  = "FERMILAB-TM-2753-AD",
    doi         = {10.48550/ARXIV.2106.03327},
    url         = {https://arxiv.org/abs/2106.03327},
    publisher   = {arXiv},
    copyright   = {Creative Commons Attribution 4.0 International}}

@phdthesis{Yuan:pyorbit,
    author      = {Yaoshuo Yuan},
    title       = {Space-charge driven transverse beam instabilities in synchrotrons},
    school      = {Technische Universit{\"a}t},
    month       = {April},
    year        = {2018},
    address     = {Darmstadt},
    url         = {http://tuprints.ulb.tu-darmstadt.de/8215/}}

@techreport{Bassetti,
    author      = "Bassetti, M and Erskine, George A",
    title       = "{Closed expression for the electrical field of a two-dimensional Gaussian charge}",
    institution = "CERN",
    address     = "Geneva, Switzerland",
    reportNumber= "CERN-ISR-TH-80-06, ISR-TH-80-06",
    year        = "1980",
    url         = "https://cds.cern.ch/record/122227"}

@book{LHCdesignreport,
    author      = "Brüning, Oliver Sim and Collier, Paul and Lebrun, P and Myers, Stephen and Ostojic, Ranko and Poole, John and Proudlock, Paul",
    title       = "{LHC Design Report}",
    publisher   = "CERN",
    address     = "Geneva, Switzerland",
    series      = "CERN Yellow Reports: Monographs",
    year        = "2004",
    url         = "https://cds.cern.ch/record/782076",
    doi         = "10.5170/CERN-2004-003-V-1"}

@inproceedings{SaaHernandez:2019ftv,
    author      = "Sa\'a Hern\'andez, Angela and Bartosik, Hannes and Biancacci, Nicolo and Hirlaender, Simon and Moreno, Daniel and Zampetakis, Michail",
    title       = {{D}etailed {C}haracterisation of the {LEIR} {I}ntensity {L}imitations for a {P}b {I}on {B}eam},
    booktitle   = {Proc. 10th International Particle Accelerator Conference (IPAC'19)},
    pages       = {3196--3199},
    paper       = {WEPTS042},
    month       = {Jun.},
    year        = {2019},
    address     = {Melbourne, Australia},
    doi         = {doi:10.18429/JACoW-IPAC2019-WEPTS042},
    url         = {http://jacow.org/ipac2019/papers/wepts042.pdf}}

@techreport{Asvesta:pyscrdt,
    author      = "Asvesta, Foteini and Bartosik, Hannes",
    title       = "{Resonance Driving Terms From Space Charge Potential}",
    institution = "CERN",
    address     = "Geneva, Switzerland",
    reportNumber= "CERN-ACC-NOTE-2019-0046",
    month       = "Oct",
    year        = "2019",
    url         = "https://cds.cern.ch/record/2696190"}

@inproceedings{Asvesta:IPAC2019,
    author      = "Asvesta, Foteini and Antoniou, Fanouria and Bartosik, Hannes and Di Giovanni, Gian Piero and Papaphilippou, Yannis",
    title       = {{C}oupling and {S}pace {C}harge {S}tudies at the {CERN} {PSB}},
    booktitle   = {Proc. 10th International Particle Accelerator Conference (IPAC'19)},
    pages       = {3192--3195},
    paper       = {WEPTS041},
    month       = {Jun.},
    year        = {2019},
    address     = {Melbourne, Australia},
    doi         = {doi:10.18429/JACoW-IPAC2019-WEPTS041},
    url         = {http://jacow.org/ipac2019/papers/wepts041.pdf}}

@techreport{IJohn,
    author      = "John, Isabelle and Bartosik, Hannes",
    title       = "{Space charge and intrabeam scattering eﬀects for Lead-ions and Oxygen-ions in the LHC injector chain}",
    institution = "CERN",
    address     = "Geneva, Switzerland",
    month       = "Jan",
    year        = "2021",
    reportNumber= "CERN-ACC-NOTE-2021-0003",
    url         = "https://cds.cern.ch/record/2749453"}

@article{Blaskiewicz:2007,
    author          = "Blaskiewicz, M. and Brennan, J. M.",
    title           = "{Bunched Beam Stochastic Cooling Simulations and Comparison with Data}",
    reportNumber    = "COOL2007-WEM2I05",
    journal         = "Conf. Proc. C",
    volume          = "07091010",
    pages           = "125--129",
    year            = "2007"}

@article{Blaskiewicz:2008,
    title           = {Operational Stochastic Cooling in the Relativistic Heavy-Ion Collider},
    author          = {Blaskiewicz, M. and Brennan, J. M. and Severino, F.},
    journal         = {Phys. Rev. Lett.},
    volume          = {100},
    issue           = {17},
    pages           = {174802},
    numpages        = {4},
    year            = {2008},
    month           = {4},
    publisher       = {American Physical Society},
    doi             = {10.1103/PhysRevLett.100.174802},
    url             = {https://link.aps.org/doi/10.1103/PhysRevLett.100.174802}}

@misc{aperture:SPS,
    title           = "{Super Proton Synchrotron Optics Repository}",
    howpublished    = {\\ \url{https://acc-models.web.cern.ch/acc-models/sps/}},
    year            = {2021}}

@misc{aperture:LEIR,
    title           = "{Low Energy Ion Ring Optics Repository}",
    howpublished    = {\\ \url{https://acc-models.web.cern.ch/acc-models/leir/}},
    year            = "2021"}

@article{Piw_BM,
    title           = {Wilson Prize article: Reflections on our experiences with developing the theory of intrabeam scattering},
    author          = {Piwinski, Anton and Bjorken, James D. and Mtingwa, Sekazi K.},
    journal         = {Phys. Rev. Accel. Beams},
    volume          = {21},
    issue           = {11},
    pages           = {114801},
    numpages        = {14},
    year            = {2018},
    month           = {11},
    publisher       = {American Physical Society},
    doi             = {10.1103/PhysRevAccelBeams.21.114801},
    url             = {https://link.aps.org/doi/10.1103/PhysRevAccelBeams.21.114801}}

@article{Martini:2016,
    author          = "Martini, Michel and Antoniou, Fanouria and Papaphilippou, Yannis",
    title           = "{Intrabeam Scattering}",
    journal         = "ICFA Beam Dyn. Newslett.",
    volume          = "69",
    pages           = "38--59",
    year            = "2016"}

\end{document}